\documentclass[letterpaper, paper, 11pt]{AAS}	
\usepackage{caption}

\usepackage{bm}
\usepackage{amsmath}
\usepackage{graphicx}
\usepackage{subcaption}
\usepackage[colorlinks=true, pdfstartview=FitV, linkcolor=black, citecolor= black, urlcolor= black]{hyperref}
\usepackage{overcite}
\usepackage{footnpag}			      	
\usepackage{float}
\usepackage{fix-cm}

\PaperNumber{26-685}

\begin{document}

\title{Estimating Harmonic Coefficients of Asteroids through LiAISON Navigation}

\author{Jun Maruyma\thanks{Graduate Research Assistant, Department of Aerospace Engineering, Iowa State University, Ames, IA},  
\ and Simone Servadio\thanks{Assistant Professor, Department of Aerospace Engineering, Iowa State University, Ames, IA}
}

\maketitle{}

\begin{abstract}
This paper evaluates the Linked Autonomous Interplanetary Satellite Orbit Navigation (LiAISON) method in near two-body asteroid systems. While LiAISON's usefulness has been proven in multibody environments such as the cislunar region, the application of a two-body system with gravity perturbations has not been sufficiently demonstrated. Vesta and Ceres are good candidates to test LiAISON, because the asteroids have sufficiently strong perturbations to use LiAISON method. The study utilizes a Multi-Model Adaptive Estimation (MMAE) framework integrated with Unscented Kalman Filters (UKF) to estimate the spacecraft states and perturbations using only range and range rate measurements. Results from the Monte Carlo analysis indicate that the proposed framework makes accurate and consistent estimates of position, velocity, and the perturbations for cases of Ceres and Vesta.        
\end{abstract}

\section{Introduction}
Recent developments in small-satellite technologies have enabled the use of CubeSats for deep-space missions. Compared with traditional spacecraft (S/C), CubeSat missions are cost-effective, easier to develop, and easier to add or remove functions~[\citen{Woellert_2011_CubeSats}]. Several recent CubeSat missions demonstrate the feasibility beyond Earth orbit. For example, the Mars Cube One (MarCO) mission, led by NASA, successfully demonstrated a prototype of a new communication and navigation system in deep space. It provided live communication for a lander mission to Mars~[\citen{Krauske_2020_Implementation}]. As a progress Artemis I deep space mission, the BioSentinel mission aims to test the effects of deep space radiation on biological systems and to demonstrate automated support for biological systems~[\citen{Tieze_2023_BioSentinel}]. These missions highlight the growing use of CubeSats across various deep-space missions.

Despite technological improvements, most deep-space missions, including CubeSat missions, rely on ground-based navigation systems, such as radiometric tracking via the Deep Space Network (DSN). Although the ground-based navigation system is effective for the recent mission, this kind of navigation system has limitations in operational cost, scalability, and the number of tracking S/Cs~[\citen{Turan_2022_Autonomous}]. In addition, the current ground-based navigation system has a time delay. Still, this delay becomes difficult to ignore when the mission duration is long or slower transponders are used~[\citen{Bertone_2018_Impact}]. Hence, there is strong motivation to develop autonomous navigation without ground-based systems, such as a geometrical approach using only angle measurements~[\citen{Choi_2026_Ambiguity}], while maintaining navigation accuracy and robustness.

One of the autonomous navigation in deep space candidates is the Linked Autonomous Interplanetary Satellite Orbit Navigation (LiAISON) method. LiAISON uses solely Satellite-to-Satellite measurements, such as Doppler range and range rate, to estimate the S/C's absolute state. As a characteristic of LiAISON, if at least one of the orbits is in an asymmetric or irregular gravitational field, this method estimates the absolute states of the S/Cs without ground support~[\citen{Hill_2007_Autonomous}]. Prior research has shown the usefulness of LiAISON in multibody and asymmetric gravity fields. Erdem et al.'s study showed that, even with less-accurate ranging measurements, this navigation method in the Earth-Moon system estimated the absolute state with sufficient accuracy~[\citen{Turan_2022_Autonomous3}]. Fujimoto et al. further extended the LiAISON concept to develop an automated gravimetry architecture for asteroids, combined with particle-based estimation methods~[\citen{Fujimoto_2016_Stereoscopic}].

However, the application of LiAISON to near two-body systems with a spherical relative gravitational field has not yet been widely researched. In the gravitational environment, due to reduction in the strength of the gravitational asymmetry raises concerns satellite-to-satellite measurements alone may not provide sufficient information to accurately estimate the gravity harmonic coefficient. This issue is linked to large asteroids or small moons with nearly spherical bodies that undergo subtle gravitational perturbations but still affect spacecraft motion.     

To use the LiAISON approach with less asymmetric orbits in near two-body systems than in multibody systems, the filter method should be more accurate than the method used in multibody systems. In current LiAISON research on multibody systems, the Extended Kalman Filter (EKF) is primarily used ~[\citen{Hill_2007_Autonomous}]~[\citen{Turan_2022_Autonomous3}]. On the other hand, some research has shown that an Unscented Kalman Filter (UKF) is particularly well-suited for relative measurement-dependent navigation systems. Lee et al. compared UKF, EKF, and the Least Squares Recursive Filter (LSRF) methods for spacecraft formation flying navigation that depends on two S/C relative measurement data and found that UKF provides much more accurate navigation than the other two filter methods~[\citen{Lee_2018_Relative}]. The research evaluating the asteroid using the Autonomous Nano-satellite Swarming (ANS) approach shows that an improved version of the UKF provides safe access to a low-altitude orbit influenced by the asteroid's strong gravity field~[\citen{Stacey_2018_Autonomous}]. Also, in the real world, the exact details of the gravitational harmonic coefficients are mostly unknown. Hence, an accurate adaptive estimation approach, such as Multi-Model Adaptive Estimation (MMAE), should be included in the system to estimate the harmonic coefficients (e.g., $J_2$ and $J_3$) and the state simultaneously. Recently, Ganganath et al. formulated a fast and accurate MMAE framework using hypothesis-diversity trigger to address misalignment estimation in star-tracker measurements~[\citen{Ganganath_2026_Compensating}].                  

Hence, this paper investigates the feasibility of estimating gravitational harmonic coefficients in near-two-body asteroid systems using LiAISON combined with MMAE and UKF. The proposed approach is evaluated using Ceres and Vesta, two large asteroids with significant gravitational perturbations. Through this research, the paper simulates observability and estimation performance using a satellite-to-satellite measurement model that closely resembles a real LiAISON scenario in near-two-body systems. The proposed methodology relies on a measurement model based solely on relative range and range-rate data between the spacecraft. State and parameter estimation are handled simultaneously within the architecture using UKF and MMAE. Additionally, a hypothesis-diversity trigger is introduced to govern the sequential refinement of gravity perturbations. 

The rest of this paper is composed as follows. The section on LiAISON for Ceres and Vesta explains why these asteroids are great candidate asteroids for testing the LiAISON approach. The LiAISON Navigation Framework section describes the dynamics models, measurement models, and the navigation architecture. Section of Numerical Simulation describes the simulation setup, results for $J_2$ coefficient estimation for Ceres and Vesta, and results for $J_2$ and $J_3$ coefficient estimation for Vesta. Finally, the Conclusion section summarizes this paper.         
 
\section{LiAISON for Ceres and Vesta}
\begin{figure}[H]
\centering
\begin{subfigure}[t]{0.48\textwidth}
\centering
\includegraphics[width=\linewidth]{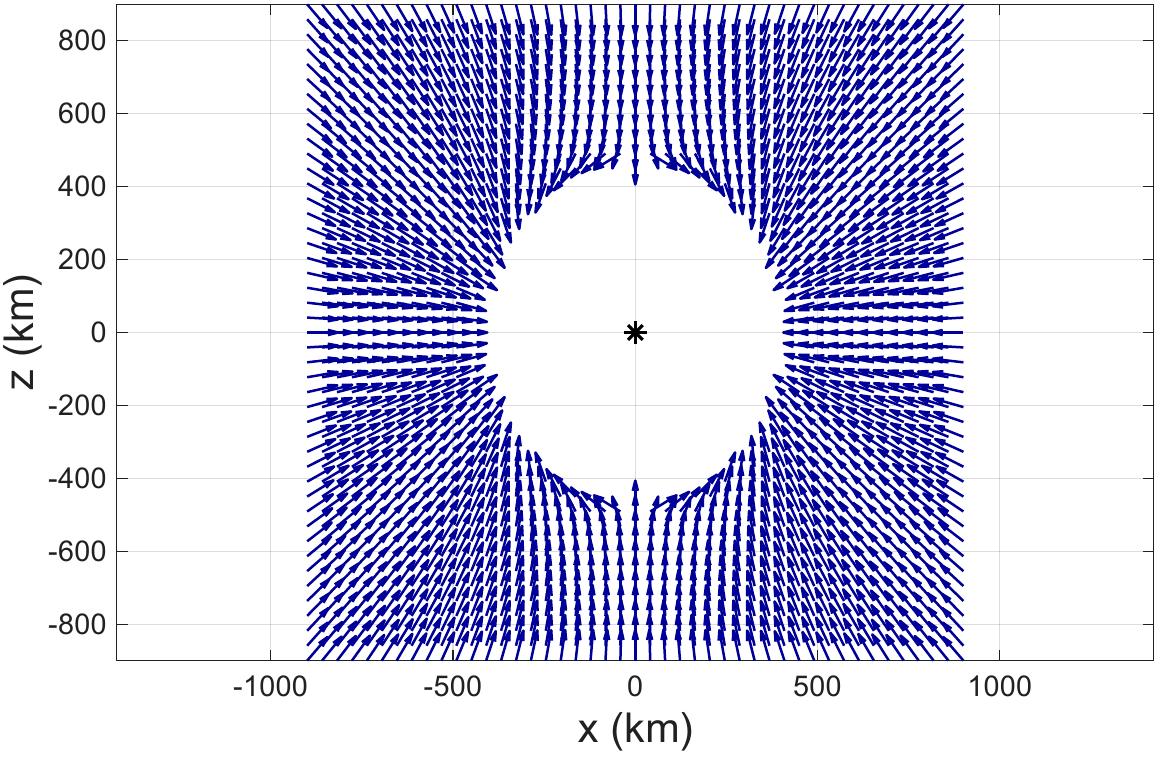}
\caption{Ceres}
\end{subfigure}
\hfill
\begin{subfigure}[t]{0.48\textwidth}
\centering
\includegraphics[width=\linewidth]{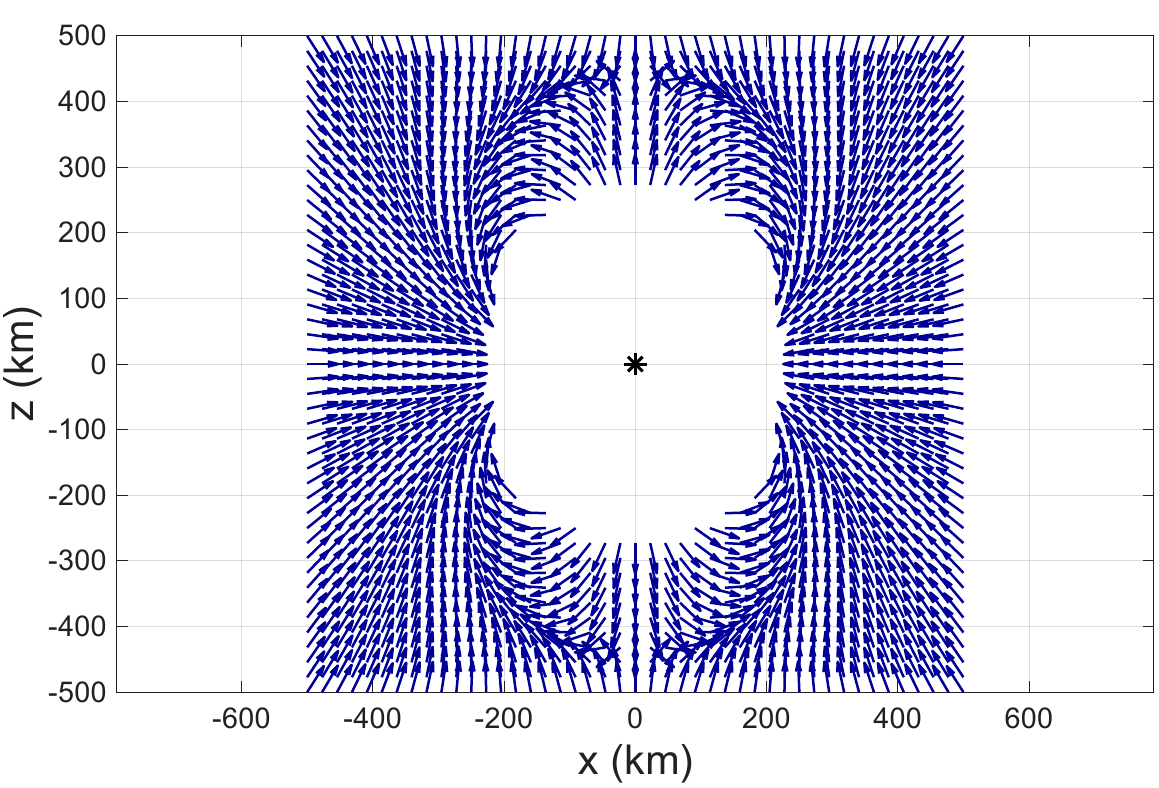}
\caption{Vesta}
\end{subfigure}
\caption{Gravitational Field From $J_2$ around Ceres (a) and Vesta (b)}
\label{fig: G_Field}
\end{figure}

\begin{figure}[htbp]
\centering
\begin{subfigure}[t]{0.48\textwidth}
\centering
\includegraphics[width=\linewidth]{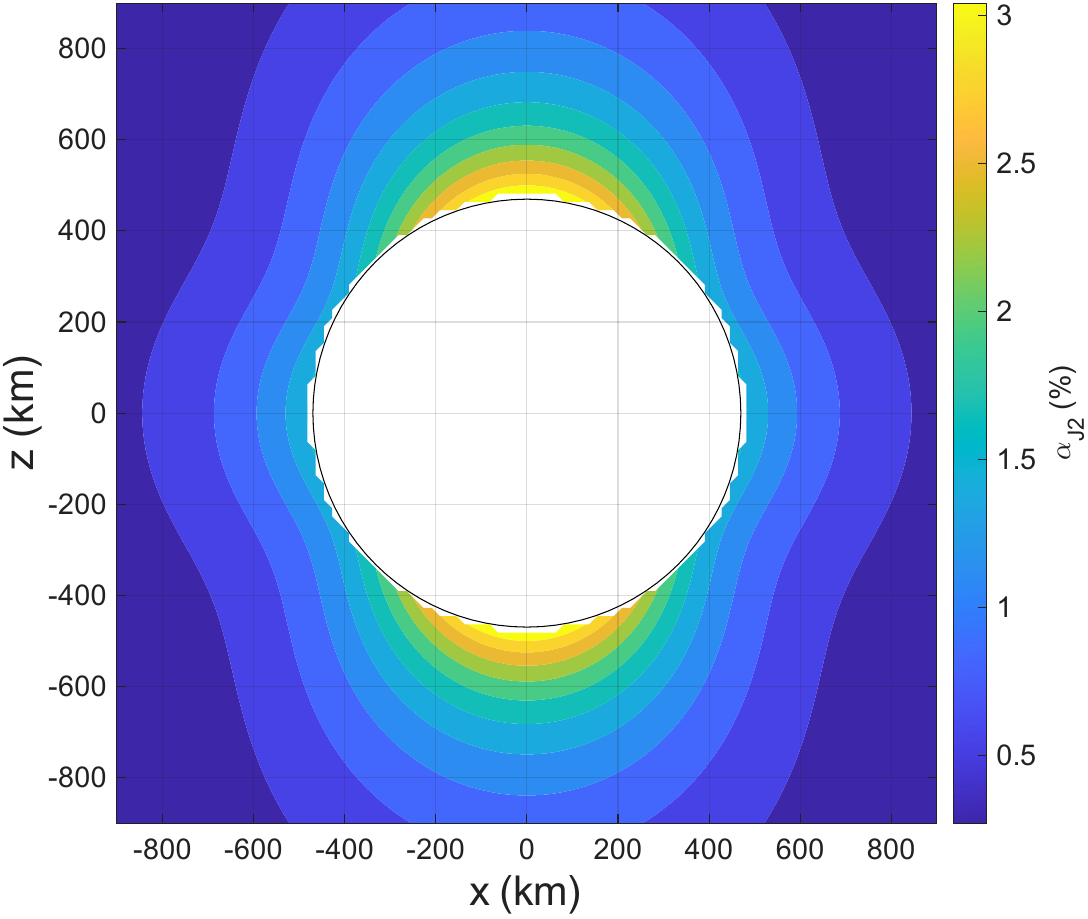}
\caption{Ceres}
\end{subfigure}
\hfill
\begin{subfigure}[t]{0.48\textwidth}
\centering
\includegraphics[width=\linewidth]{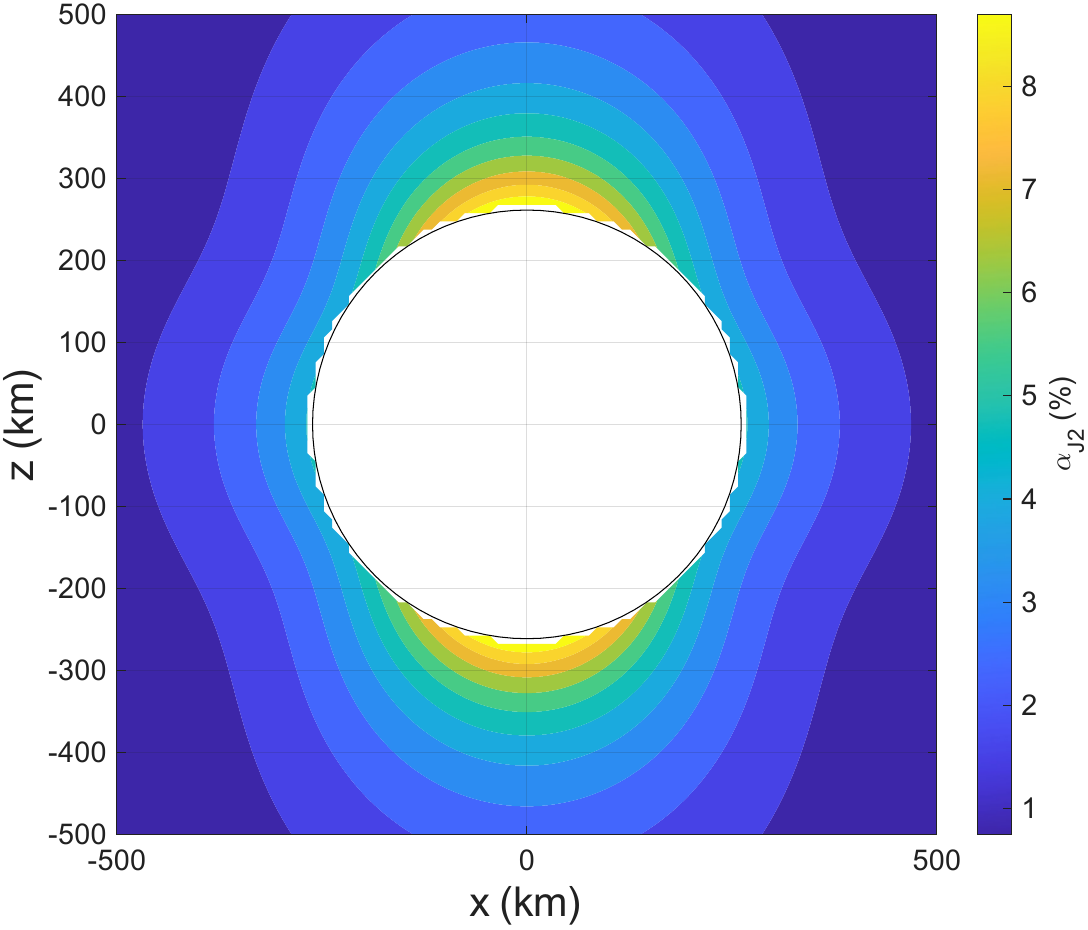}
\caption{Vesta}
\end{subfigure}
\caption{Map of $\alpha$ for acceleration due to $J_2$ in Ceres (a) and Vesta (b)}
\label{fig: A_Field}
\end{figure}

This research evaluates the performance of the LiAISON method for near two-body systems, using range and range-rate measurements without ground observations. As a requirement of this method, the states of spacecraft (S/C) must always be observable from relative measurements between each S/C. Also, the LiAISON technique requires that one of the S/Cs has a unique orbit in an asymmetric gravity field of sufficient strength. Currently, many studies focus on the application of LiAISON to the Cislunar three-body problem~[\citen{Hill_2007_Autonomous},~\citen{Turan_2022_Autonomous3}]. Although Earth and Moon do not have large $J_2$ or other perturbations that individually cause enough asymmetric gravity fields~[\citen{Curtis_2014_Orbital}], these gravity fields interact, creating an asymmetric field region in the Earth-Moon system. For the two-body system, a planet or an asteroid should have a large enough gravity perturbation to satisfy the prerequisite of LiAISON. Among the planets and asteroids for which accurate gravitational harmonic coefficient data have been obtained in previous studies, Ceres and Vesta, which are located in the main asteroid belt between Mars and Jupiter~[\citen{NASA_JPLC},~\citen{NASA_JPLV}], are good candidates. It is because these asteroids have sufficiently strong $J_2$ harmonic coefficients that produce an asymmetric gravity field without $J_3$ influence or other perturbations. The gravitational field models of Ceres and Vesta, influenced by the $J_2$ coefficient, are shown in Figure~\ref{fig: G_Field}. To improve the visibility of asymmetry, the $J_2$ coefficient influence is exaggerated in Figure~\ref{fig: G_Field}.
 In this model, both asteroids' gravitational fields, especially Vesta's, become asymmetric unless the x and z axes are used as the reference lines. Hence, for the aspect of asymmetry, both asteroids seem to satisfy the prerequisite of the LiAISON method. However, for the practical purpose, the relative strength of the asymmetric acceleration must be greater than the observation and process noises. To assess the field's strength, the parameter $\alpha$ is calculated by using the formula~[\citen{Hill_2007_Autonomous}]:  
\begin{equation}
  \alpha_{j}(x,y,z)=\frac{|a_{j}(x,y,z)|}{\sum_{i=1}^{n}|a_{i}(x,y,z)|}
\end{equation}
where $a_j$is each point of the gravitational acceleration influenced by the gravitational harmonic coefficients. The denominator of this equation represents the sum of each point of the gravitational acceleration. If using this $\alpha$ equation for every area around the asteroids, we can visualize the relative strength of acceleration from $J_2$ coefficient influence, such as Figure~\ref {fig: A_Field}.

In Figure~\ref{fig: A_Field}, the gravitational coefficient effects are not exaggerated. The $\alpha$ values for Vesta are larger than the values for Ceres. However, this $\alpha$ value is sufficiently larger than the typical range and range-rate measurement model errors that are used in many LiAISON method-related research without $J_3$ and other perturbations. Hence,  Ceres and Vesta are great candidates for evaluating the Performance of absolute-state determination using the LiAISON approach in near two-body systems.     

\section{LiAISON Navigation Framework}
This section explains the dynamics and measurement models, as well as the LiAISON-based autonomous navigation framework, in two-body asteroid systems with perturbations. As a proposed methodology, satellite-to-satellite measurements with UKF and MMAE architectures are used to estimate the spacecraft (S/C) state and gravitational harmonic coefficients without a ground-based tracking system reference.                

\subsection{Dynamics Model}
 A two-body system is used to model each S/C's motion when assuming the asteroid as the center of gravity and the S/C as the gravitational disturbance. Let $\mathbf{r}$ and $\mathbf{v}$ be the position and velocity vector of the S/C in the asteroid-centered inertial frame, respectively. The differentiation of the position and velocity is given by:
\begin{equation}
    \dot{\mathbf{r}} = \mathbf{v}, \quad \dot{\mathbf{v}} = -\frac{\mu}{r^3}\mathbf{r} + \mathbf{a}_{J_n}(\mathbf{r})
\end{equation}
where $\mu$ is the asteroid's gravitational parameter and $\mathbf{a}_{J_n}(\mathbf{r})$ represents the acceleration from the zonal gravitational harmonics influence. If the spin axis of a massive asteroid is the same as the axis of rotational symmetry of the gravitational field and the asteroid shape is assumed to be an oblate spheroid, the gravitational potential equations with rotational symmetric perturbation become~[\citen{Curtis_2014_Orbital}]~[\citen{Battin_1999_An}]:
\begin{equation}
    \begin{gathered}
    V(r, \phi) = -\frac{\mu}{r} + \Phi(r, \phi) \\
    \Phi(r, \phi) = \frac{\mu}{r}\sum_{k=2}^{\infty} J_k \left(\frac{R}{r}\right)^k P_k(\cos \phi)
    \end{gathered}
\end{equation}
where $\phi$ is the polar angle from the radial z-axis of the asteroid's frame, $R$ is the equatorial radius of the asteroid, and $P_k$ is the Legendre polynomials. $\Phi(r,\phi)$ for $J_{2}$ coefficient becomes:
\begin{equation}
    \Phi_{J_2}(r,\phi)=\frac{J_{2}}{2}\frac{\mu}{r}(\frac{R}{r})^{2}(3\cos^{2}\phi-1)
\end{equation}
The acceleration from the $J_{2}$ coefficient is described as:
\begin{equation}
    \mathbf{a}_{J_2}(\mathbf{r}) = -\nabla \Phi_{J_2} = -\frac{\partial\Phi_{J_2}}{\partial x}\hat{\mathbf{i}} - \frac{\partial\Phi_{J_2}}{\partial y}\hat{\mathbf{j}} - \frac{\partial\Phi_{J_2}}{\partial z}\hat{\mathbf{k}}
\end{equation}
If we differentiate the rotational symmetric harmonic by each positional state of x, y, and z, the gradient of the perturbing potential formula from the $J_2$ coefficient is:
\begin{equation}
     \begin{gathered}
       \frac{\partial\Phi_{J_2}}{\partial x}=-\frac{3}{2}J_{2}\frac{\mu}{r^{2}}\left(\frac{R}{r}\right)^{2}\frac{x}{r}\left[5\left(\frac{z}{r}\right)^{2}-1\right] \\
     \frac{\partial\Phi_{J_2}}{\partial y}=-\frac{3}{2}J_{2}\frac{\mu}{r^{2}}\left(\frac{R}{r}\right)^{2}\frac{y}{r}\left[5\left(\frac{z}{r}\right)^{2}-1\right] \\
     \frac{\partial\Phi_{J_2}}{\partial z}=-\frac{3}{2}J_{2}\frac{\mu}{r^{2}}\left(\frac{R}{r}\right)^{2}\frac{z}{r}\left[5\left(\frac{z}{r}\right)^{2}-3\right]
     \end{gathered}
\end{equation}
 $\Phi(r,\phi)$ for $J_{3}$ coefficient becomes: 
 \begin{equation}
    \Phi_{J_3}(r,\phi)=\frac{J_{3}}{2}\frac{\mu}{r}(\frac{R}{r})^{3}(5\cos^3\phi -3\cos\phi)
\end{equation}
Similar to the acceleration from $J_2$ coefficient, $J_3$ harmonic coefficient acceleration can be computed by differentiation of $\Phi_{J_3}(r,\phi)$ by each positional state:  
\begin{equation}
     \begin{gathered}
     \mathbf{a}_{J_3}(\mathbf{r}) = -\nabla \Phi_{J_3} = -\frac{\partial\Phi_{J_3}}{\partial x}\hat{\mathbf{i}} - \frac{\partial\Phi_{J_3}}{\partial y}\hat{\mathbf{j}} - \frac{\partial\Phi_{J_3}}{\partial z}\hat{\mathbf{k}} \\
      \frac{\partial\Phi_{J_3}}{\partial x}=-\frac{5}{2}J_{3}\frac{\mu}{r^{2}}\left(\frac{R}{r}\right)^{3}\frac{x}{r}\frac{z}{r}\left[7\left(\frac{z}{r}\right)^{2}-3\right] \\
     \frac{\partial\Phi_{J_3}}{\partial y}=-\frac{5}{2}J_{3}\frac{\mu}{r^{2}}\left(\frac{R}{r}\right)^{3}\frac{y}{r}\frac{z}{r}\left[7\left(\frac{z}{r}\right)^{2}-3\right] \\
     \frac{\partial\Phi_{J_3}}{\partial z}=-\frac{1}{2}J_{3}\frac{\mu}{r^{2}}\left(\frac{R}{r}\right)^{3}\left[35(\frac{z}{r})^{4}-30\left(\frac{z}{r}\right)^{2}+3\right]
     \end{gathered}
\end{equation}
If assuming that unmodeled acceleration from other perturbations, solar radiations, and other factors can be modeled as a continuous white noise with spectral density, $\sigma_a^2$, the process noise matrix $Q$ can be defined as: 
\begin{equation}
    \mathbf{Q}=\mathbf{blkdiag}(\mathbf{Q_b},\mathbf{Q_b}),
    \quad \mathbf{Q_b}=\sigma_a^2 \begin{bmatrix} \frac{\Delta t^3}{3}\mathbf{I_3} &\frac{\Delta t^2}{2}\mathbf{I_3} \\ \frac{\Delta t^2}{2}\mathbf{I_3} & \Delta t\mathbf{I_3} \end{bmatrix}
\end{equation}
where $Q_b$ is each S/C process noise covariance matrix from the unmodeled acceleration assumption. In this $Q$ model, we assumed that the process noise matrix $Q$ is kept smaller than the influence from the gravitational harmonic coefficients of $J_2$ and $J_3$, which are used to distinguish perturbation acceleration from unmodeled acceleration. If including the acceleration from $J_2$, $J_3$, and unmodeled factors, the differentiation of the position and velocity becomes: 
\begin{equation}
    \label{eqn: Dynamics}
    \dot{\mathbf{r}} = \mathbf{v}, 
    \quad \dot{\mathbf{v}} = -\frac{\mu}{r^3}\mathbf{r} + \mathbf{a}_{J_2}(\mathbf{r}) + \mathbf{a}_{J_3}(\mathbf{r})+\mathbf{w},  \quad
    \mathbf{w}\sim N(0,\mathbf{Q})
\end{equation}

\subsection{Measurement Model}
The navigation framework depends on the satellite-to-satellite measurements. In this model, there are two kinds of observations: relative range ($\rho$) and relative range rate ($\dot{\rho}$). The estimated state vectors for the two spacecraft's position and velocity are:
\begin{equation}
    \mathbf{x}=[\mathbf{r}_1, \mathbf{v}_1, \mathbf{r}_2, \mathbf{v}_2]^T
\end{equation}
where $r_1$ and $v_1$ are corresponding to the first satellite and $r_2$ and $v_2$ are corresponding to the second satellite. 
The measurement vector $\mathbf{z}$ is defined as:
\begin{equation}
  \begin{gathered}
    \mathbf{z} = \begin{bmatrix} \rho \\ \dot{\rho} \end{bmatrix} + \mathbf{\nu},
    \quad  \mathbf{\nu} \sim N(0,\mathbf{R}), \quad \mathbf{R}=\mathbf{diag} (\sigma_{\rho}^2,\sigma_{\dot{\rho}}^2)\\
     \rho = \|\mathbf{r}_1 - \mathbf{r}_2\|\\
     \dot{\rho} = \frac{(\mathbf{r}_1 - \mathbf{r}_2) \cdot (\mathbf{v}_1 - \mathbf{v}_2)}{\rho}
  \end{gathered}
\end{equation}
where $\nu$ is the Gaussian measurement noise generated from the noise covariance $\mathbf{R}$ with zero mean, and $\sigma_{\rho}^2$ and $\sigma_{\dot{\rho}}^2$ are the variances of the relative range and relative range rate. In this measurement model, the external measurement model is not used to assess the usefulness of the LiAISON system based solely on inter-satellite measurement data. Also, we assume that this measurement model can be tracked perfectly when S/Cs are observable to each other.   

\subsection{UKF with LiAISON}
The UKF is selected to estimate the absolute state of a spacecraft (S/C) orbiting in a nonlinear two-body system with gravitational perturbations. The UKF is much more accurate at assessing a nonlinear system than the EKF. It is because the UKF uses the parameters of $\alpha$, $\beta$, and $k$ to control the spread of sigma points. It is necessary for computing the posterior mean and covariance to at least the second order of the Taylor series expansion, surpassing the linear approximation in EKF~[\citen{Wan_2000_The}]. In addition, the UKF does not need to use the Jacobian matrix, which is a large matrix composed of the derivatives of the S/C dynamics and often causes estimation issues, unlike other Kalman filters, such as the EKF. The UKF algorithm utilized in this paper follows the formulation described in~[\citen{Wan_2000_The}] and~[\citen{Candan_2025_Adaptive}],
as summarized in equations~\ref{eqn: lambda} to~\ref{eqn: Pk}. The sigma-point weights are determined by the parameters $\alpha$, $\beta$, and $k$. From $\alpha$, $\beta$, and $\kappa$, the scale parameter, $\lambda$, is computed as:
\begin{equation}
    \label{eqn: lambda}
    \lambda = \alpha^2(L + \kappa) - L
\end{equation}
where $L=12$ is the dimension of S/Cs' state vectors. The weight and the covariance weight are computed from the sigma points. The formula for this calculation becomes: 
\begin{equation}
    W_{m}^{(0)} = \frac{\lambda}{L+\lambda}
\end{equation}
\begin{equation}
     W_{c}^{(0)} = \frac{\lambda}{L+\lambda} + (1-\alpha^{2}+\beta) \\
\end{equation}
\begin{equation}
    W_{m}^{(i)} = W_{c}^{(i)} = \frac{1}{2(L+\lambda)}, \quad i=1,...,2L
\end{equation}
As a prediction step, the sigma points and the uncertainty regions are calculated by using the previous state estimation and covariance:
\begin{equation}
    \mathbf{\chi}_{k-1}^{(i)} = 
    \begin{cases}
    \hat{\mathbf{x}}_{k-1} & \text{for } i = 0 \\
    \hat{\mathbf{x}}_{k-1} + (\sqrt{(L+\lambda)\mathbf{P}_{k-1}})_i & \text{for } i = 1, \dots, L \\
    \hat{\mathbf{x}}_{k-1} - (\sqrt{(L+\lambda)\mathbf{P}_{k-1}})_{i-L} & \text{for } i = L+1, \dots, 2L
    \end{cases}
\end{equation}
These sigma points are computed using the nonlinear transition function that explains orbit physics. The propagated formula becomes: 
\begin{equation}
    \mathbf{\chi}_{k|k-1}^{(i)} = f(\mathbf{\chi}_{k-1}^{(i)},\mathbf{u}_{k-1})
\end{equation}
After the propagation of the sigma points, the average of this propagation is used to find new predicted positions and new uncertainty by using the formulas:  
\begin{equation}
    \mathbf{\hat{x}}_{k|k-1}=\Sigma_{i=0}^{2L}W_{m}^{(i)}\mathbf{\chi}_{k|k-1}^{(i)}
\end{equation}
\begin{equation}
    \mathbf{P}_{k|k-1}=\Sigma_{i=0}^{2L}W_{c}^{(i)}(\mathbf{\chi}_{k|k-1}^{(i)}-\mathbf{\hat{x}}_{k|k-1})(\mathbf{\chi}_{k|k-1}^{(i)}-\mathbf{\hat{x}}_{k|k-1})^{T}+\mathbf{Q}
\end{equation}
Since this navigation system measures the relative range and range rate of S/Cs, the measurement data is necessary to convert the S/Cs' absolute state data. This translation works by using the formula:
\begin{equation}
    \mathbf{\gamma}_{k}^{i} = h(\mathbf{\chi}_{k|k-1}^{i})
\end{equation}
By calculating the predicted measurement and measurement covariance, the cross covariance, and the Kalman gain are computed:
\begin{equation}
    \mathbf{\hat{{z}}}_k = \sum_{i=0}^{2L} W_m^{(i)} \mathbf{\gamma}_k^{(i)}
\end{equation}
\begin{equation}
    \mathbf{{S}}_k = \sum_{i=0}^{2L} W_c^{(i)} (\mathbf{\gamma}_k^{(i)} - \mathbf{\hat{{z}}}_k) (\mathbf{\gamma}_k^{(i)} - \mathbf{\hat{{z}}}_k)^T + \mathbf{R}
\end{equation}
\begin{equation}
    \mathbf{{T}}_k = \sum_{i=0}^{2L} W_c^{(i)} (\mathbf{\chi}_{k|k-1}^{(i)} - \mathbf{\hat{{x}}}_{k|k-1}) (\mathbf{\gamma}_k^{(i)} - \mathbf{\hat{{z}}}_k)^T
\end{equation}
\begin{equation}
    \mathbf{{K}}_k = \mathbf{{T}}_k\mathbf{{S}}_k^{-1}
\end{equation}
From the Kalman gain and the prediction calculation results, the estimation of the state and covariance is updated by utilizing the equations:
\begin{equation}
    \mathbf{\hat{x}}_{k} = \mathbf{\hat{x}}_{k|k-1} + \mathbf{K}_{k}(\mathbf{z}_{k} - \mathbf{\hat{z}}_{k})
\end{equation}
\begin{equation}
    \label{eqn: Pk}
    \mathbf{P}_{k} = \mathbf{P}_{k|k-1} - \mathbf{K}_{k}\mathbf{S}_{k}\mathbf{K}_{k}^{T}
\end{equation}

\subsection{Multi-Model Adaptive Estimation}
\begin{figure}[htbp]
\centering
\includegraphics[width=0.9\linewidth]{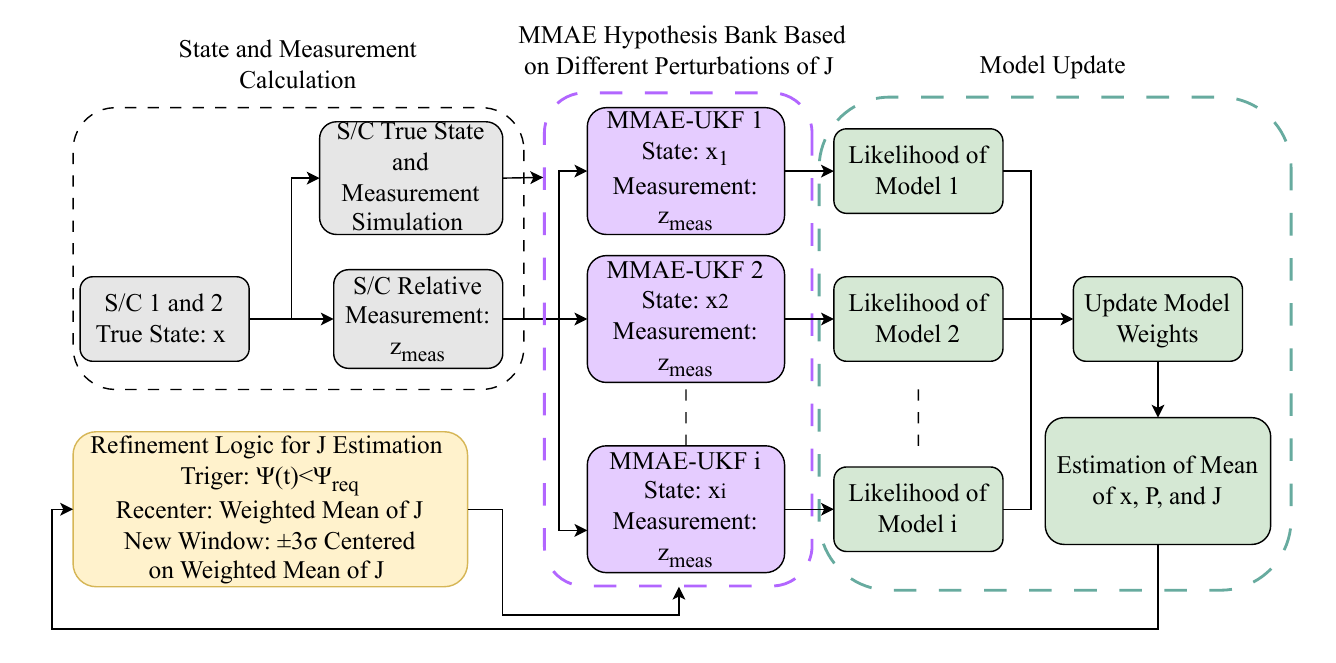}
\caption{LiAISON Combined with MMAE and UKF Architecture. The diagram illustrates the main phase of the proposed framework: State and Measurement Calculation, MMAE Hypothesis Bank, Model Update, and Refinement Logic.}
\label{fig: MMAEUKF}
\end{figure}

The Multi-Model Adaptive Estimation (MMAE) framework estimates gravitational harmonic coefficients without extending the spacecraft (S/C) state vector while simultaneously predicting the S/C states. A discrete set of candidate gravitational models is defined, each corresponding to a different assumed value of the $J_{2}$ and $J_{3}$ coefficients, based on the true state and measurement calculations. For each candidate model, the UKF computes the state estimate and the associated covariance, as explained in the previous section. At each measurement update, the likelihood of the observed measurement residual is computed for each model using its covariance. These likelihoods are used to update the model's probability weights using a Bayesian weighting scheme. To calculate the likelihood of the probability of getting measurements, the formula is used~[\citen{Ganganath_2026_Compensating}]:
\begin{equation}
    \mathcal{L}_{i,k} = \frac{1}{\sqrt{(2\pi)^m |\mathbf{S}_{i,k}|}} \exp \left( -\frac{1}{2} \mathbf{e}_{i,k}^{T} \mathbf{S}_{i,k}^{-1} \mathbf{e}_{i,k} \right)
\end{equation}
where $\mathbf{e}$ is the innovation, and S is the covariance related to the state propagation that shows the confidence of the current prediction from each UKF. After getting all the models' weight and state propagation data, the weight, $w_{i}$, for this estimation framework is updated by utilizing Bayes’ rule equation: 
\begin{equation}
    w_{i,k} = \frac{\mathcal{L}_{i,k} w_{i,k-1}}{\sum_{i=1}^M \mathcal{L}_{i,k} w_{i,k-1}}
\end{equation}
From the formula, the maximum weight model is determined by the measurement error. The new state vector, state covariance, and each harmonic coefficient $J_l$ are determined from the calculated weight. 
\begin{equation}
    \mathbf{\hat{x}}_{MMSE, k} = E[x_k|Z_k] = \sum_{i=1}^{M}   w_{i,k} \mathbf{\hat{x}}_{i,k}
\end{equation}
\begin{equation}
    \mathbf{P}_{MMSE, k} = \sum_{i=1}^{M} w_{i,k} \left( \mathbf{P}_{i,k} + [\mathbf{\hat{x}}_{i,k} - \mathbf{\hat{x}}_{MMSE, k}][\mathbf{\hat{x}}_{i,k} - \mathbf{\hat{x}}_{MMSE, k}]^T \right)
\end{equation}
\begin{equation}
    \hat{J}_{l,MMSE,k} =
    \sum_{i=1}^{M}w_{i,k}\hat{J}_{l,i,k}
\end{equation}
The hypothesis diversity trigger, $\Psi(t)$, is used to determine whether the filter successfully detects the localized solution. The trigger, $\Psi(t)$, is defined as~[\citen{Ganganath_2026_Compensating}]: 
\begin{equation}
    \Psi(t)=\frac{100A_{t}}{N}, \quad A_{t}=\left({\sum_{i=1}^M w_{i,t}^2}\right)^{-1}
\end{equation}
where $N$ is the number of models in the MMAE parameter grid, $w_{i}$ is a normalized weight, and $A_{t}$ is the inverse of the weight concentration. The large number of $\Psi(t)$ represents that the models are uniformly distributed, while the small $\Psi (t)$ is a sign of the determination of localized solutions. When $\Psi (t)$ is small enough, the grid center becomes the current weighted mean of the harmonic coefficients, and the grid range is set to three times the standard deviation of the weighted harmonic coefficient models. At the same time, each model weight $w_{i}$ is initialized so that $\Psi (t)$ equals 100, indicating equal distribution. Figure~\ref{fig: MMAEUKF} summarizes the proposed gravitational perturbation estimation method using LiAISON with MMAE and UKF.  

\section{Numerical Simulation}
This section mainly evaluates the proposed autonomous navigation framework by modeling two S/Cs orbiting at low altitudes around Ceres and Vesta. To evaluate the framework, we simulate two scenarios. The first scenario focuses on estimating the $J_2$ gravitational coefficient for Ceres and Vesta. The second scenario focuses on estimating the $J_2$ and $J_3$ coefficients for Vesta. To assess the statistical performance and robustness of this approach, Monte Carlo analyses are conducted for each scenario. 

\subsection{Simulation Setup}
\begin{figure}[htbp]
\centering
\begin{subfigure}[t]{0.48\textwidth}
\centering
\includegraphics[width=\linewidth]{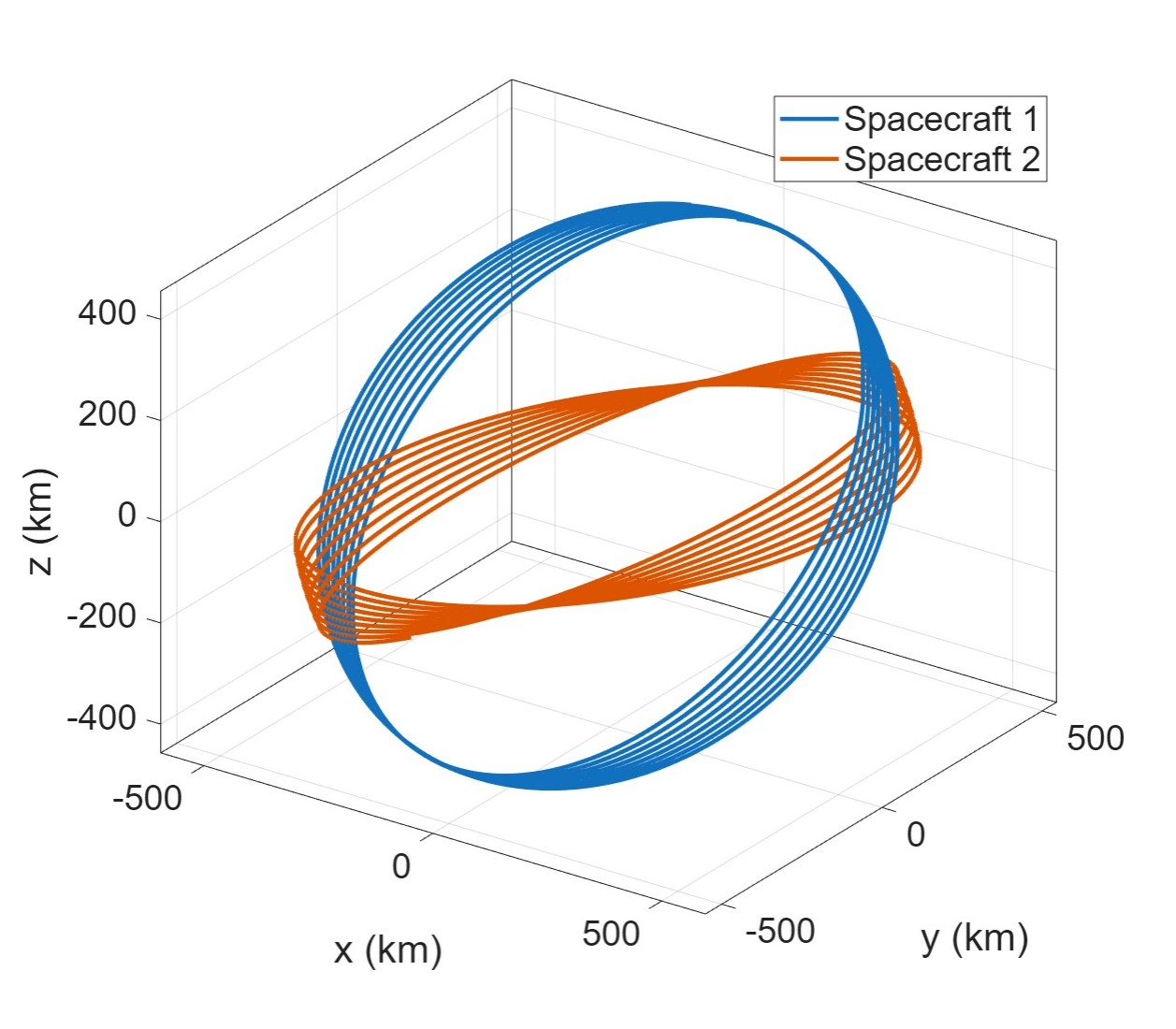}
\caption{Ceres}
\end{subfigure}
\hfill
\begin{subfigure}[t]{0.48\textwidth}
\centering
\includegraphics[width=\linewidth]{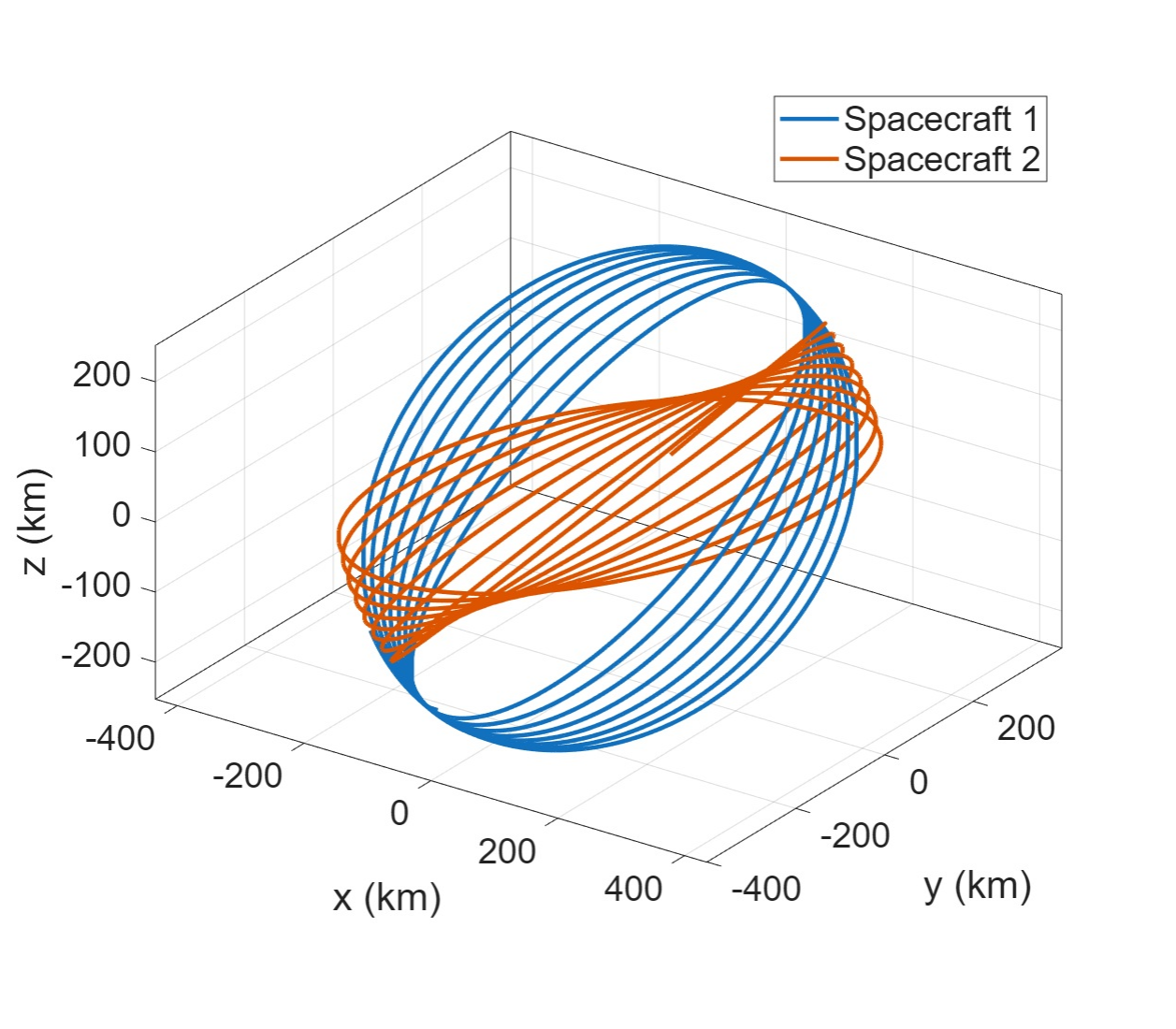}
\caption{Vesta}
\end{subfigure}
\caption{Two Spacecraft Orbits around Ceres(a) and Vesta(b) Influenced by Only $J_2$}
\label{fig: AsetroidP}
\end{figure}

The mission scenario consists of two spacecraft (S/C) orbiting at low altitudes around Ceres and Vesta. In scenario 1, the only $J_2$ coefficients for Ceres and Vesta are estimated with each S/C state prediction. In scenario 2, we tried to estimate $J_2$ and $J_3$ coefficients for Vesta simultaneously. The fundamental orbital information for each asteroid is listed in Table~\ref{tab: OrbitP}, including the gravitational parameter, the mean radius, and the true $J_2$ and $J_3$ coefficients~[\citen{NASA_JPLC},~\citen{NASA_JPLV}]~[\citen{Konopliv_2018_The},~\citen{Konopliv_2014_The}]. S/C 1 for each asteroid has the same orbital parameters written in Table~\ref{tab: OrbitP}. The initial true anomaly of each S/C is randomly selected between $0^{\circ}$ and $359^{\circ}$ as a uniform distribution selection. For simulation purposes, before any calculation, these orbital parameters are converted to Cartesian coordinates. As an example of S/C trajectories for the simulations, Figure~\ref{fig: AsetroidP} shows trajectories of two S/Cs around Ceres and Vesta for eight periodic times of S/C 1 ($2T$), influenced only by the $J_2$ coefficients.

\begin{table}[htbp]
	\fontsize{11}{11}\selectfont
    \caption{Initial Orbital Parameters}
   \label{tab: OrbitP}
        \centering 
   \begin{tabular}{c | c | c } 
      \hline 
                                              & Ceres     & Vesta\\
      \hline 
      Semi-Major Axis for S/C 1 and 2 ($km$)                  & 777.778, 777.778    & 388.889, 388.889 \\
      Eccentricity for S/C 1 and 2                            & 0.2857, 0.2857    & 0.2857, 0.2857 \\
      Inclination for S/C 1 and 2  ($^{\circ}$)                & 45, 135        & 45, 135 \\
      RAAN for S/C 1 and 2 ($^{\circ}$)                       & 0, 45         &  0, 45\\
      Argument of Periapsis for S/C 1 and 2  ($^{\circ}$)      & 0, 0         &  0, 0\\ 
      Mean Radius ($km$)                      & 469.7     & 261.385\\
      Gravitational Parameter ($km^3/s^2$)    & 62.6284   & 17.2883\\
      $J_2$, $J_3$                                    & 0.011851, $-4.15241\cdot10^{-5}$  & 0.031779, $-3.31055\cdot10^{-3}$\\
      \hline
   \end{tabular}
\end{table}

When concentrating on Figure~\ref{fig: AsetroidP}, the trajectories of two S/Cs around both asteroids are not periodic since the gravitational perturbation introduces non-periodic shifts in the shape and position of the orbits. In addition, this influence is increasing over time. When comparing orbits of S/Cs around Ceres and Vesta, the trajectory of Vesta is more unstable because of a larger $J_2$ value. Additionally, with $J_3$ coefficient influence, each S/C trajectory becomes more complicated.          

The filtering parameters, including variance related to the initial state covariance $P_0$ and variance linked to the process and measurement noise covariances $Q$ and $R$, are shown in Table~\ref{tab: ErrorC}.

\begin{table}[htbp]
    \fontsize{11}{11}\selectfont
    \caption{Variance of Initial State and Noise Covariances}
    \label{tab: ErrorC}
    \centering 
    \begin{tabular}{c | c | c } 
      \hline 
      $P_0$ & Position ($\sigma_r^2$): $3(0.2)^2$ $km^2$ & Velocity ($\sigma_v^2$): $3(0.05\times10^{-3})^2$ $km^2/s^2$\\
      \hline
      $Q$   & \multicolumn{2}{c}{Spectral Density ($\sigma_a^2$): $(1.0\cdot10^{-12})^2$ $km^2/s^4$} \\ 
      \hline 
      $R$   & Range ($\sigma_{\rho}^2$): $0.1(0.01)^2$ $km^2$ & Range Rate ($\sigma_{\dot{\rho}}^2$): $0.1(0.001)^2$ $km^2/s^2$  \\
      \hline
    \end{tabular}
\end{table}
Before our navigation framework takes over, initial coarse orbital state information is already provided by standard deep-space navigation. Consequently, the initial covariance values, $P_0$, are larger than those typically seen in relative measurement-based navigation.

We used the MMAE framework with UKF to estimate the gravitational coefficients, with UKF parameters set to $\alpha=0.01$, $\beta=2$, and $\kappa=0$. A block of UKF is executed in parallel as one hypothesis. All filter has the same measurement model. For each hypothesis, the stacked innovation value provides a likelihood of updating each model weight $w_i$. The 10 percent threshold for $\Psi (t)$ that triggers the refinement has been chosen. 

The performance of the LiAISON method with the MMAE and the UKF framework is evaluated across both scenarios using 100 Monte Carlo analyses ($N_{MC}=100$). Each run is initialized with the same conditions, including the initial state covariance, $P_0$, and noise covariance $Q$ and $R$. To assess the filter's Performance, the Root Mean Squared Error (RMSE) $\Xi$ for each S/C's position $r$, velocity $v$, and each gravitational harmonic coefficient $J_l$ is calculated. The formulas are defined as~[\citen{Michaelson_2024_Particle}]: 
\begin{equation}
    \Xi_{r,k} = \sqrt{\frac{1}{N_{MC}}\sum_{i=1}^{N_{MC}}(r_k^{i}-\widehat{r_k^i})(r_K^{i}-\widehat{r_k^i})^T}
\end{equation}
\begin{equation}
    \Xi_{v,k} = \sqrt{\frac{1}{N_{MC}}\sum_{i=1}^{N_{MC}}(v_k^{i}-\widehat{v_k^i})(v_k^{i}-\widehat{v_k^i})^T}
\end{equation}
\begin{equation}
    \Xi_{J_{l,k}} = \sqrt{\frac{1}{N_{MC}}\sum_{i=1}^{N_{MC}}(J_{l.k}^{i}-\widehat{J_{l,k}^i})(J_{l,k}^{i}-\widehat{J_{l,k}^i})^T}
\end{equation}

\subsection{Simulation Results for \texorpdfstring{$J_2$}{J2} Estimation for Ceres and Vesta}
In the first simulation case, we analyze the proposed framework for estimating the $J_2$ gravitational harmonic coefficient for Ceres and Vesta. In this scenario, the true and estimated dynamics are identical and include normal, $J_2$, and process-noise-influenced acceleration components. The simulation duration for this scenario is set to two periodic times of S/C 1 ($2T$) for each asteroid with 100 time steps and $\Delta t=\frac{2T}{100}$. The initial $J_2$ value grid for both asteroids is defined as the minimum value of $10^{-3}$ and the maximum value of $10^{-1}$, while the grid should have 200 candidate values for the true coefficient to be almost the center of the grid. 

\begin{figure}[htbp]
\centering
\begin{subfigure}[t]{0.49\textwidth}
\centering
\includegraphics[width=\linewidth]{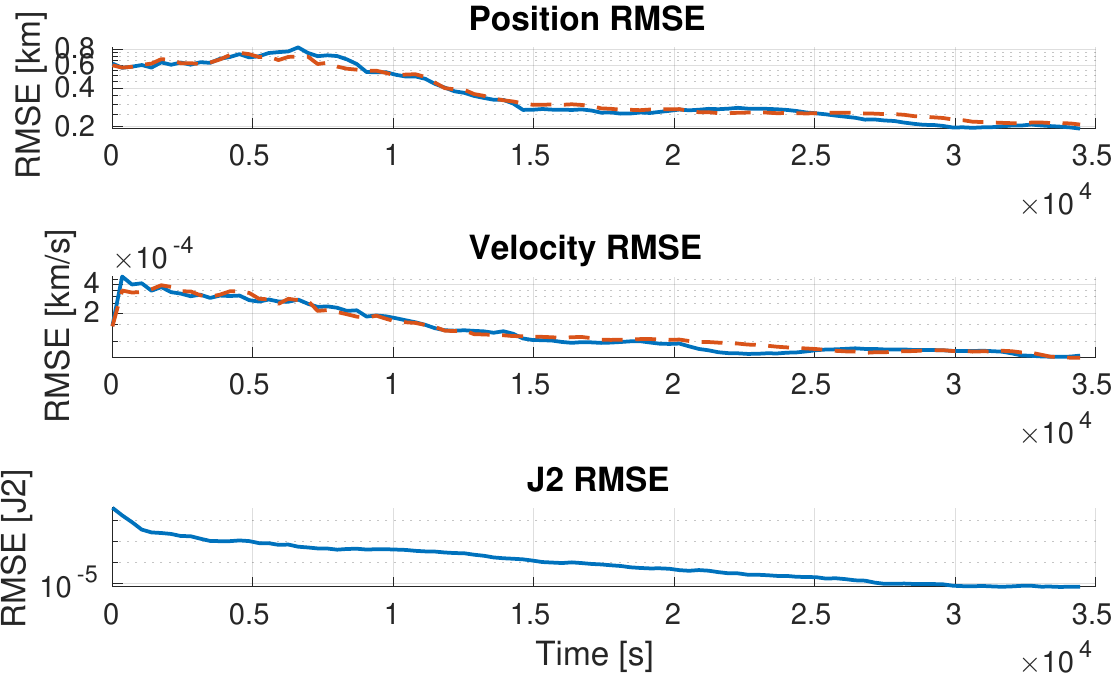}
\caption{Ceres}
\label{fig: RMSEC}
\end{subfigure}
\hfill
\begin{subfigure}[t]{0.49\textwidth}
\centering
\includegraphics[width=\linewidth]{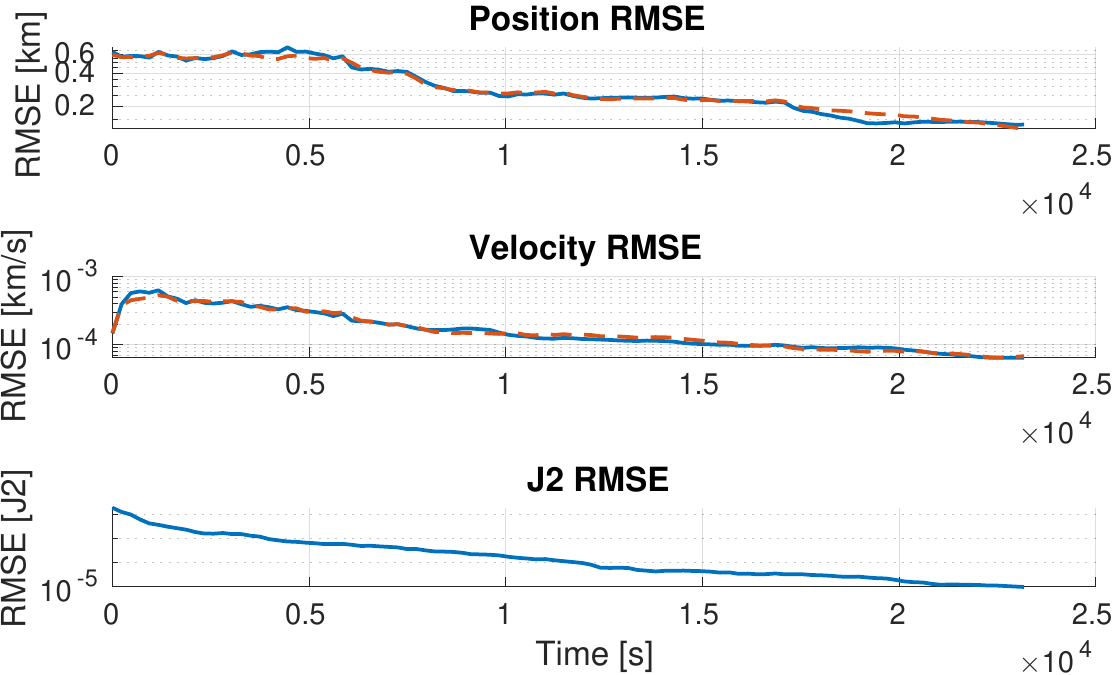}
\caption{Vesta}
\label{fig: RMSEV}
\end{subfigure}
\caption{RMSE of S/C 1 and S/C 2 for Ceres(a) and Vesta(b). Top: Position RMSE for S/C 1 (blue) and S/C 2 (orange). Middle: Velocity RMSE S/C 1 (blue) and S/C 2 (orange). Bottom: $J_2$ RMSE}
\label{fig: RMSE}
\end{figure}

The statistical performance of the proposed framework for this scenario is evaluated using a 100-run Monte Carlo analysis. As a simulation result, a nominal convergence rate of about 90$\%$ is achieved for Vesta, while that for Ceres is about 80$\%$. Other trials are recognized as outliers and excluded from the final result calculations. It is because their steady-state errors slightly exceed the 3$\sigma$ range and cause the results trends to become slightly unclear, although these excluded runs remain stable and bounded. Figure~\ref{fig: RMSE} shows the RMSE for each S/C state and $J_2$ coefficients for both asteroids.     

According to Figure~\ref{fig: RMSE}, the position and velocity RMSEs for each S/C in both asteroids gradually decrease to small values with some oscillation. This oscillation suggests that a geometric relationship between two S/Cs makes it difficult to estimate the absolute states, even though the S/Cs are observable in this scenario. When comparing the RMSEs for the positions of two asteroids and accounting for the percentage of outliers, the position RMSE for Ceres slightly struggles to converge compared with Vesta's because smaller gravitational perturbations make state estimation less stable. However, overall, the LiAISON estimation architecture maintains stable state estimation for both spacecrafts around the orbits of the target asteroids, as evidenced by the convergence of the position and velocity RMSEs for Ceres and Vesta.           

\begin{figure}[htbp]
\centering
\begin{subfigure}[t]{0.47\textwidth}
    \centering
    \includegraphics[width=\linewidth]{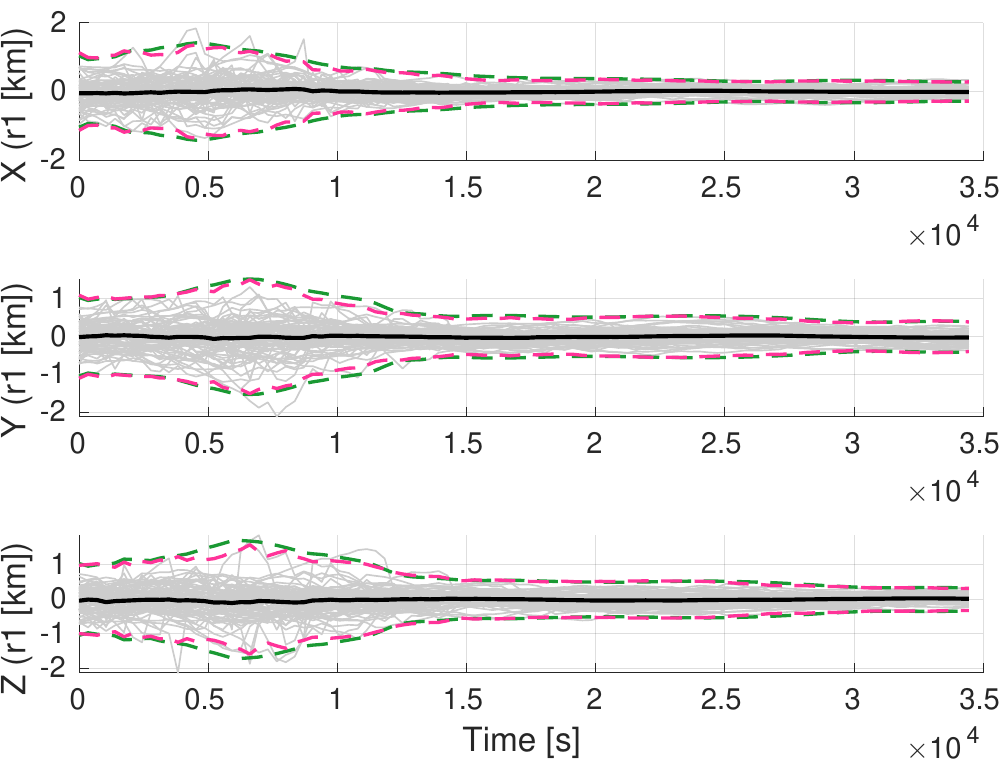}
    \caption{Spacecraft 1 Position}
\end{subfigure}
\hfill 
\begin{subfigure}[t]{0.47\textwidth}
    \centering
    \includegraphics[width=\linewidth]{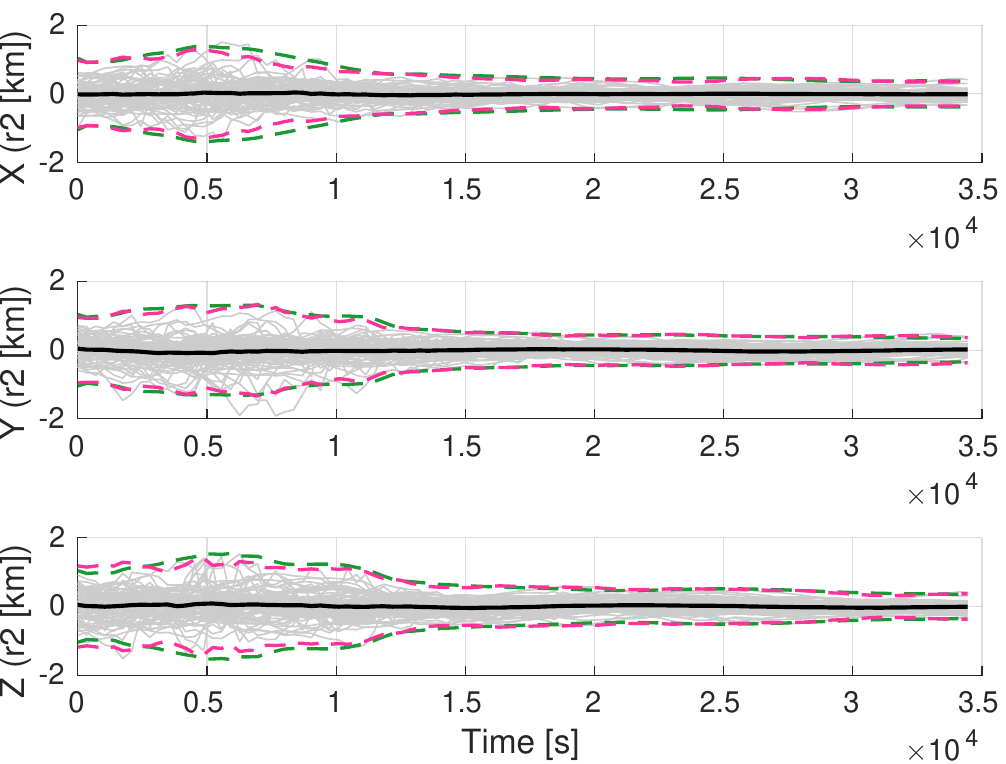}
    \caption{Spacecraft 2 Position}
\end{subfigure}
\\[2ex] 
\begin{subfigure}[t]{0.47\textwidth}
    \centering
    \includegraphics[width=\linewidth]{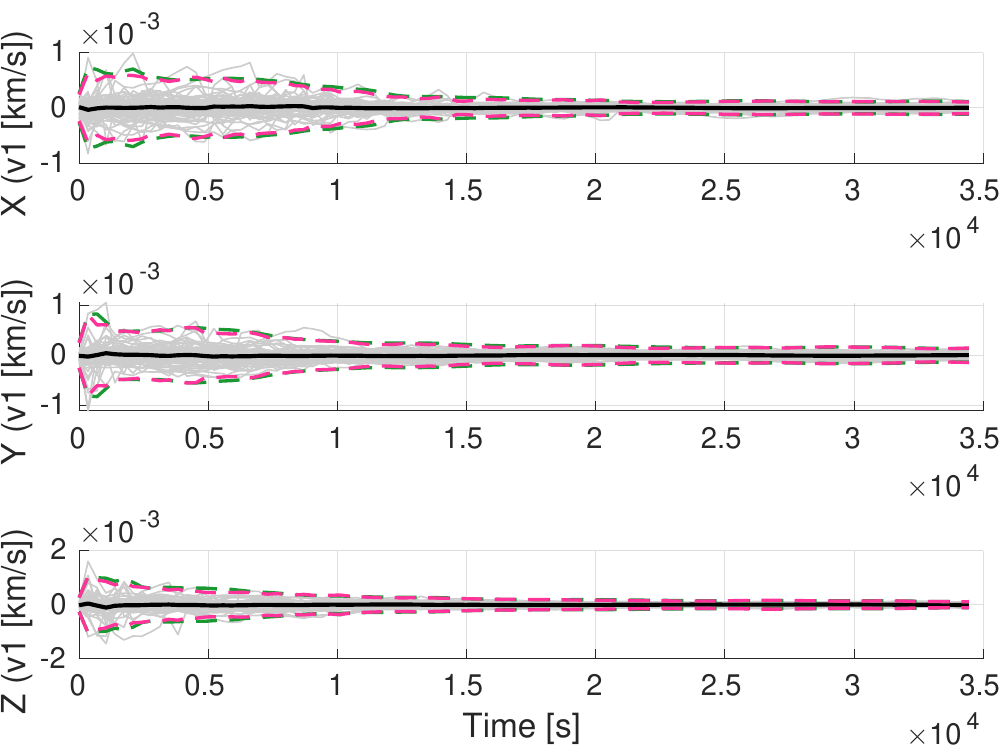}
    \caption{Spacecraft 1 Velocity}
\end{subfigure}
\hfill
\begin{subfigure}[t]{0.47\textwidth}
    \centering
    \includegraphics[width=\linewidth]{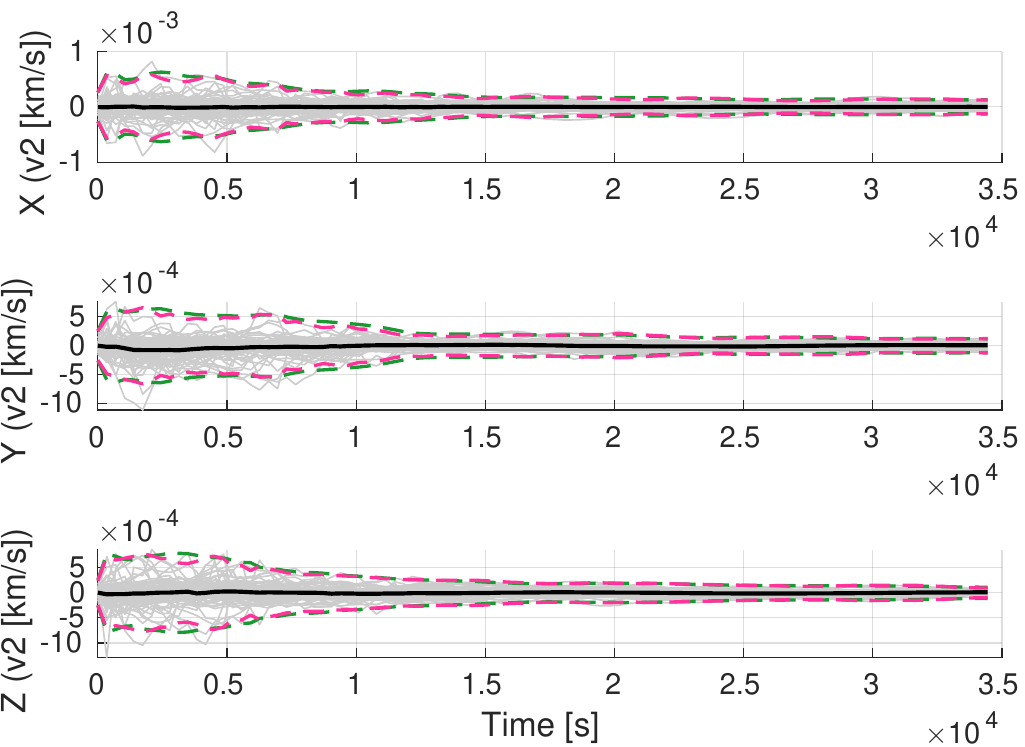}
    \caption{Spacecraft 2 Velocity}
\end{subfigure}
\caption{Two S/Cs Position and Velocity Consistency for Ceres. All plots include estimated $\pm3\sigma$ (green) and empirical $\pm3\sigma$ (pink) bounds across Monte Carlo trials (gray).}
\label{fig: ConsistC}
\end{figure}

\begin{figure}[htbp]
\centering
\begin{subfigure}[t]{0.47\textwidth}
    \centering
    \includegraphics[width=\linewidth]{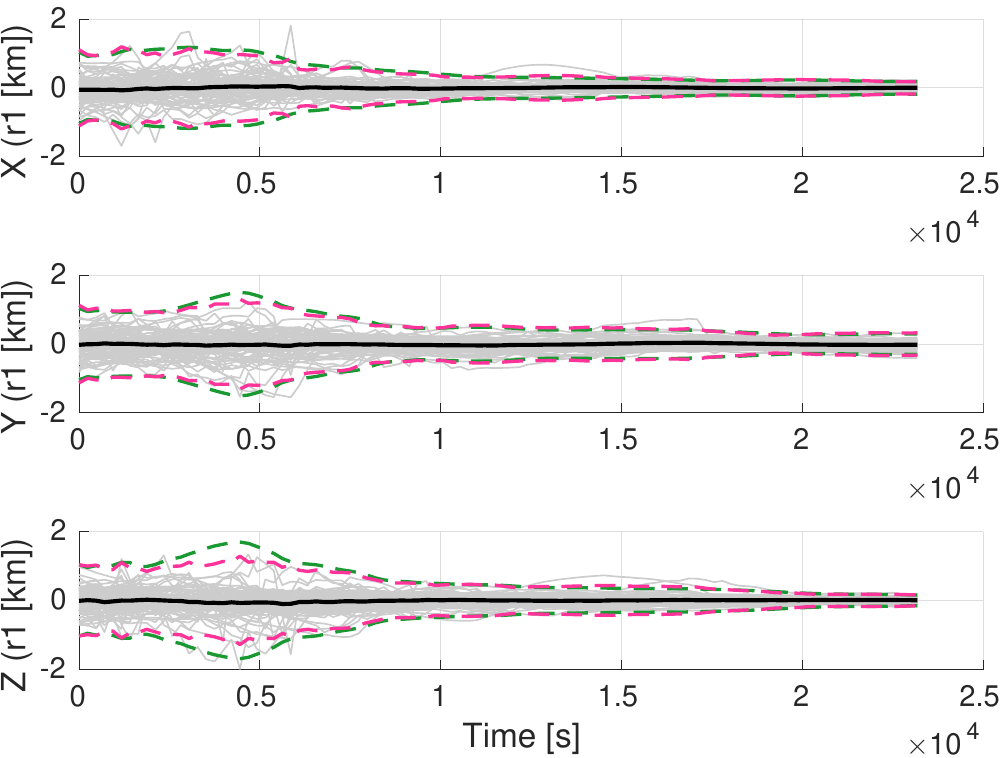}
    \caption{S/C 1 Position}
\end{subfigure}
\hfill
\begin{subfigure}[t]{0.47\textwidth}
    \centering
    \includegraphics[width=\linewidth]{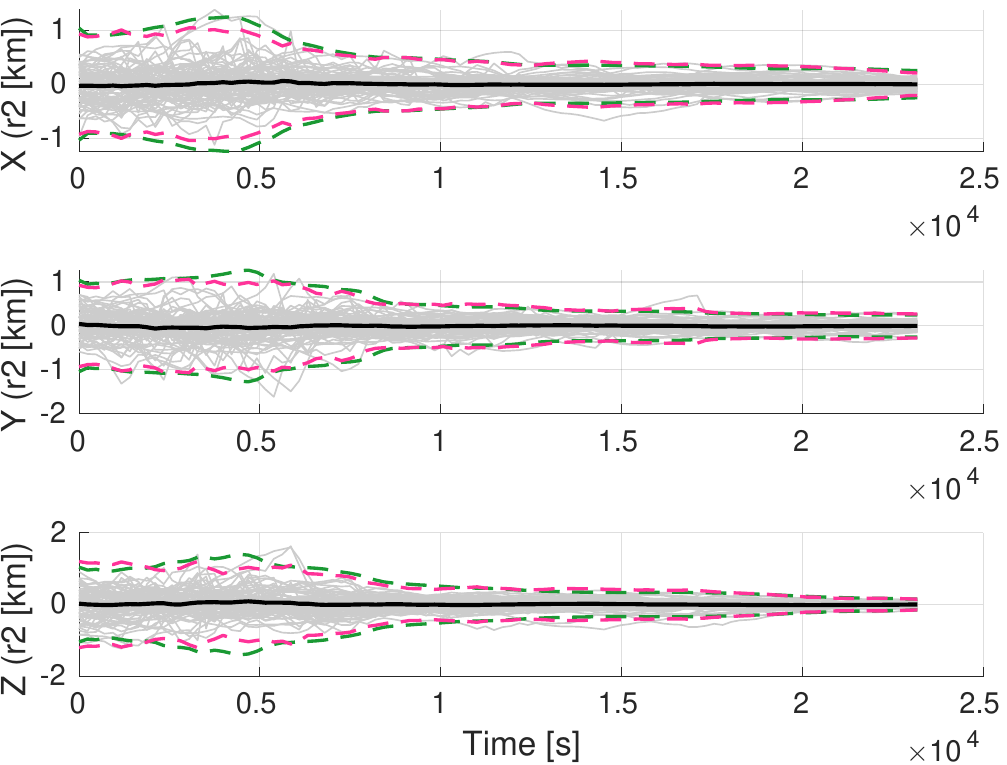}
    \caption{S/C 2 Position}
\end{subfigure}
\\[2ex]
\begin{subfigure}[t]{0.47\textwidth}
    \centering
    \includegraphics[width=\linewidth]{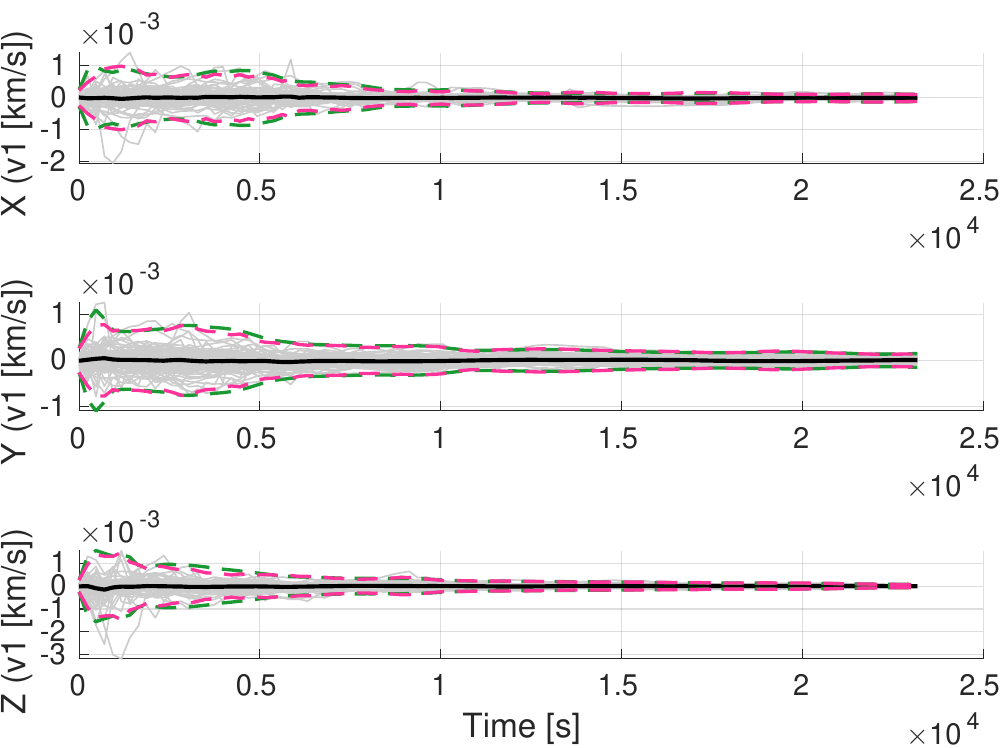}
    \caption{S/C 1 Velocity}
\end{subfigure}
\hfill
\begin{subfigure}[t]{0.47\textwidth}
    \centering
    \includegraphics[width=\linewidth]{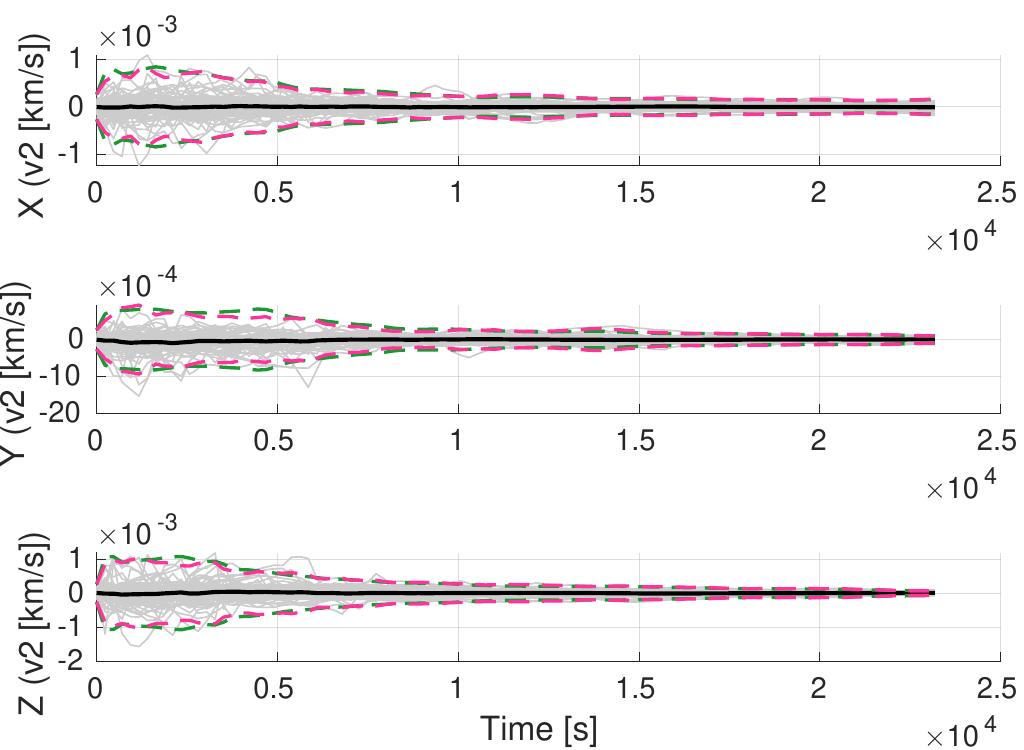}
    \caption{S/C 2 Velocity}
\end{subfigure}
\caption{Two S/Cs Position Consistency for Vesta. All plots include estimated $\pm3\sigma$ (green) and empirical $\pm3\sigma$ (pink) bounds across Monte Carlo trials (gray).}
\label{fig: ConsistV}
\end{figure}

\begin{figure}[htbp] 
\centering
\begin{subfigure}[t]{0.48\textwidth}
\centering
\includegraphics[width=\linewidth]{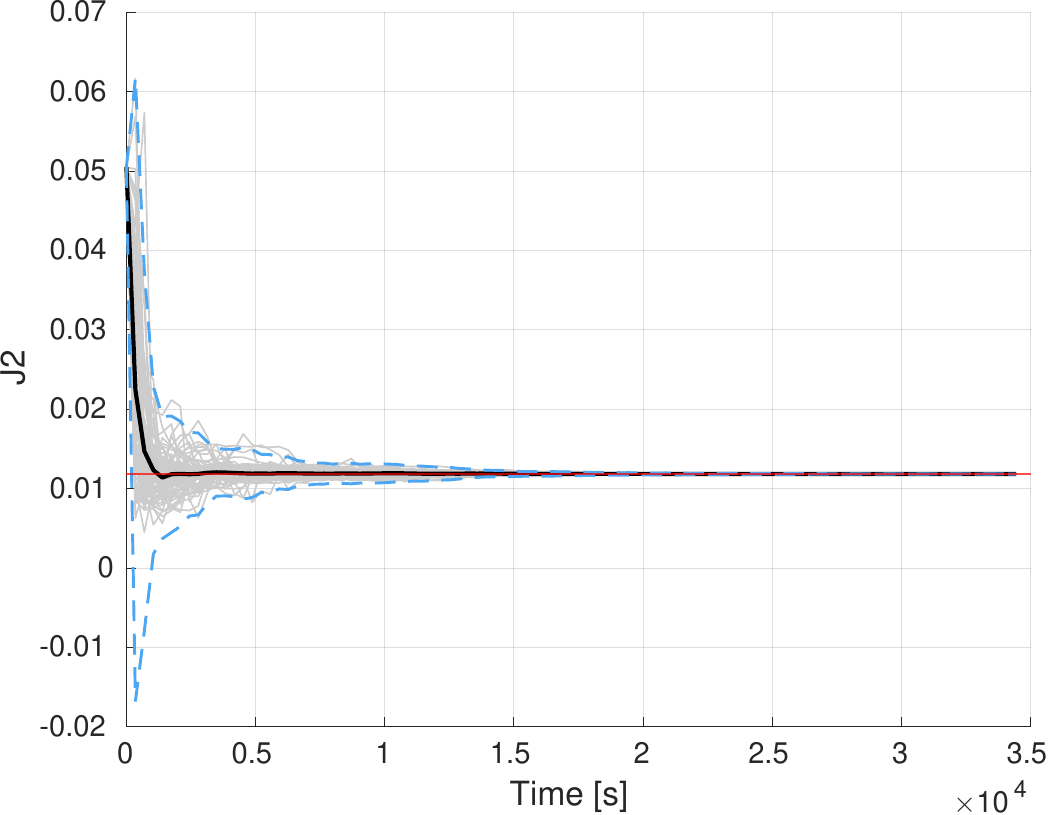}
\caption{Ceres}
\end{subfigure}
\hfill
\begin{subfigure}[t]{0.48\textwidth}
\centering
\includegraphics[width=\linewidth]{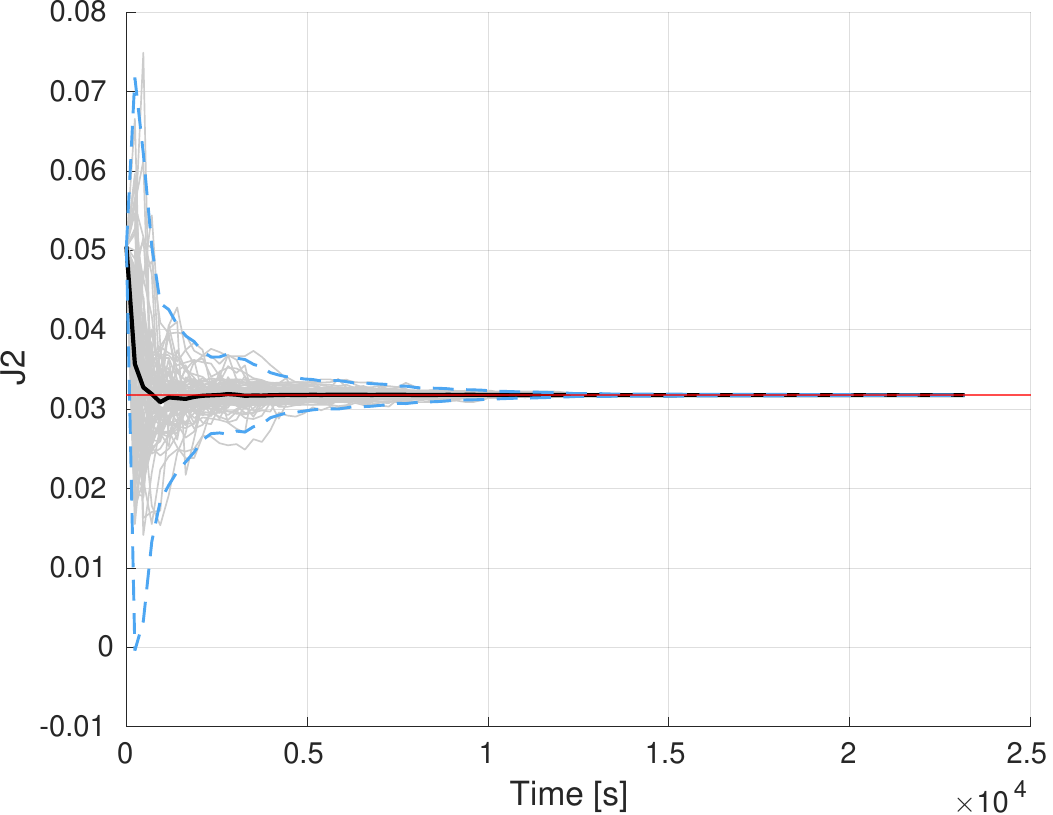}
\caption{Vesta}
\end{subfigure}
\caption{$J_2$ Estimation Consistency of Ceres(a) and Vesta(b) with $\pm3\sigma$ (blue) bounds}
\label{fig: ConsistJ}
\end{figure}

For additional validation, the filter consistency analysis is also performed for the position, velocity, and $J_2$ coefficient in both asteroid situations. Figures~\ref{fig: ConsistC} and~\ref{fig: ConsistV} represent the consistency of the position and velocity components of each S/C around Ceres and Vesta. Figure~\ref{fig: ConsistJ} shows the estimation consistency of the $J_2$ coefficient for the situations of Ceres and Vesta.

For the position-velocity consistency figures for spacecraft in both scenarios, most estimation errors in each Monte Carlo run are within the mean and the estimated consistency covariance. The mean of errors converges to almost zero from the initial simulation time. If including most outlier results, they are so close to the 3$\sigma$ range. The results confirm that the filters on Ceres and Vesta provide consistent and reliable state estimation. Also, $J_2$ mean values in both asteroids converge to true values rapidly, and each Monte Carlo run's mean values are within the covariance. Hence, the MMAE with the UKF framework provides the fast, consistent, and reliable estimation of the $J_2$ coefficient.

\begin{figure}[htbp] 
\centering
\begin{subfigure}[t]{0.49\textwidth}
\centering
\includegraphics[width=\linewidth]{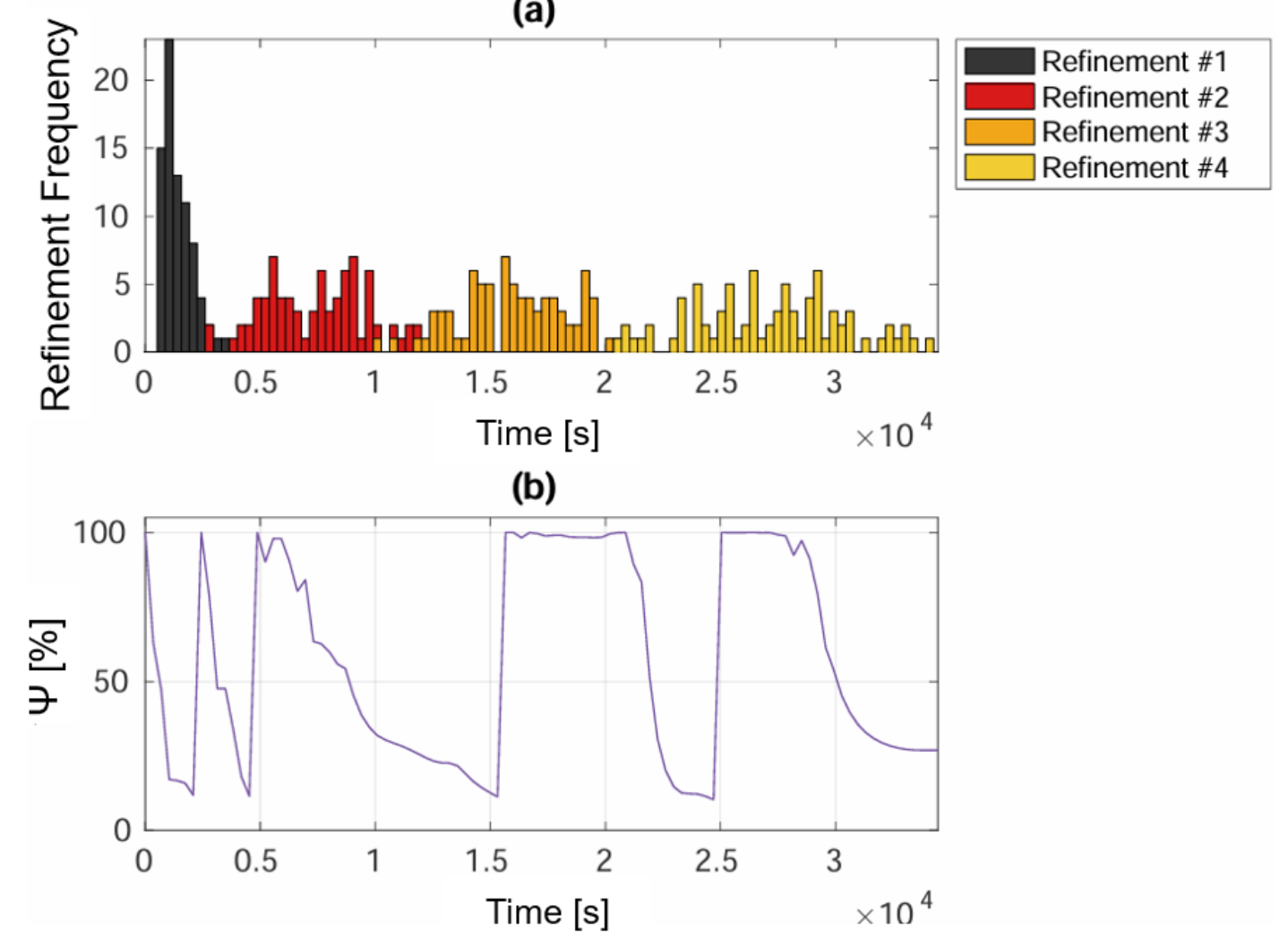}
\caption*{Ceres}
\end{subfigure}
\hfill
\begin{subfigure}[t]{0.49\textwidth}
\centering
\includegraphics[width=\linewidth]{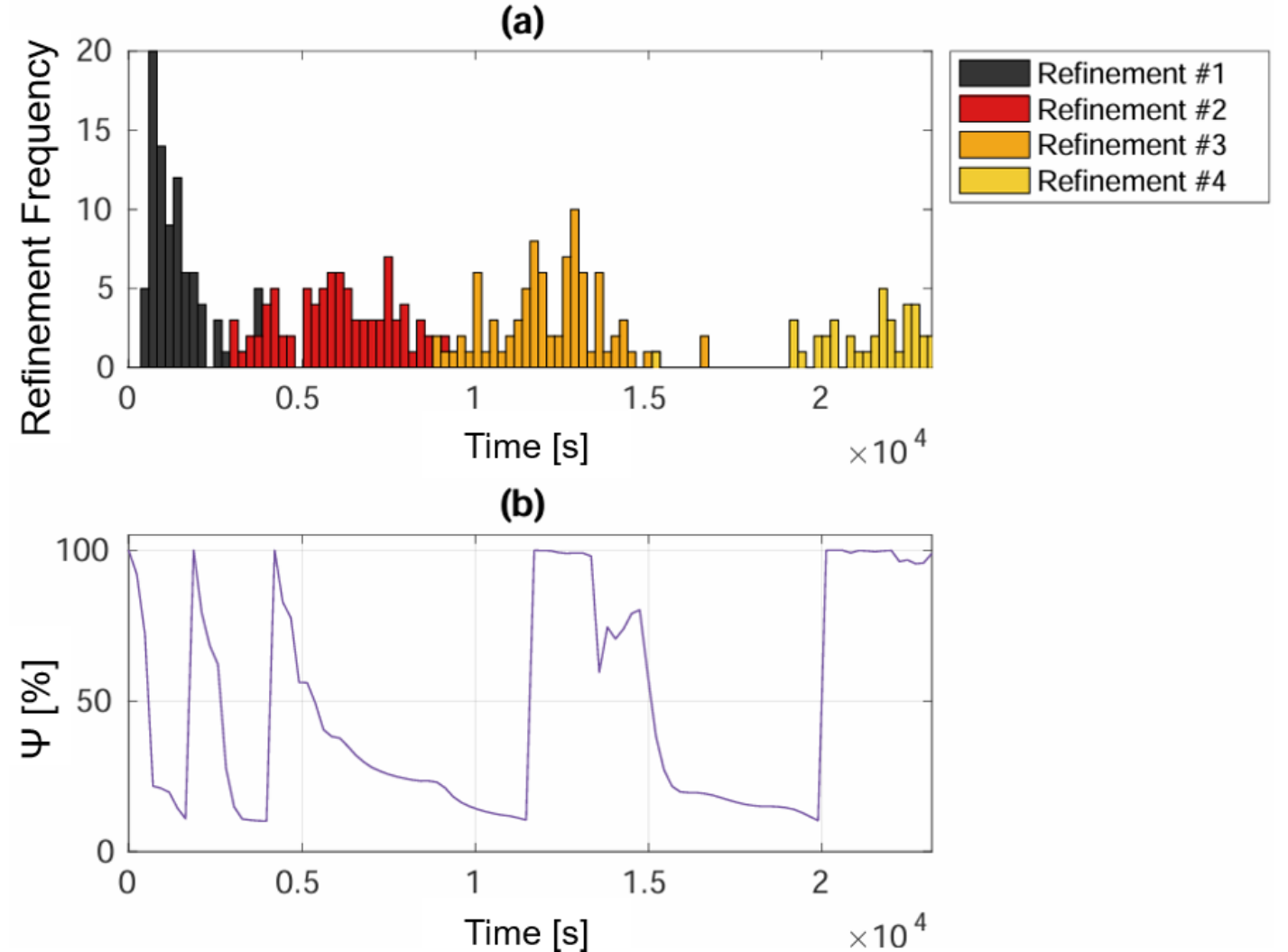}
\caption*{Vesta}
\end{subfigure}
\caption{Histogram of Refinement Event Times over 100 Monte Carlo Runs (a) and Example Evolution of Hypothesis Diversity Metric $\Psi(t)$ (b) for Ceres and Vesta}
\label{fig: J2psi}
\end{figure}

The behaves of the MMAE $\Psi(t)$ hypothesis trigger strategy for Ceres and Vesta is represented in Figures~\ref{fig: J2psi}. The upper figure shows the histogram of refinement event times across 100 Monte Carlo analyses with each refinement table. The lower figure displays the example time evolution of the $\Psi (t)$ trigger.

The evolution behavior of $\Psi (t)$ in Ceres and Vesta indicates that the MMAE $\Psi (t)$ trigger approach does not converge during the simulation. However, the RMSE and Monte Carlo consistency figures indicate it has almost converged. As a possible reason, this MMAE approach estimates $ J_2$ in detail. However, the $J_2$ coefficient is accurate enough to generate a reasonable S/C state compared with the true state. This suggests that the MMAE $\Psi (t)$ trigger approach is necessary to include a strategy for stopping the estimation when it is sufficiently close to a desired value, especially in the small-fraction estimation scenario. 

Therefore, the simulation results demonstrate that the LiAISON approach produces an accurate state estimate with sufficient consistency for the near two-body systems of Ceres and Vesta. Also, the LiAISON MMAE with the UKF framework consistently estimates reasonable $J_2$ values.

\subsection{Simulation Results for \texorpdfstring{$J_2$}{J2} and \texorpdfstring{$J_3$}{J3} Estimation for Vesta}
In the second simulation case, the suggested method is evaluated by simultaneously estimating multiple coefficients. In this scenario, the true and estimated dynamics models are identical to the model written in equation~\ref{eqn: Dynamics}. This scenario's simulation time is set to 4 periodic times of S/C 1 ($4T$) around Vesta, with 200 time steps and $\Delta t = \frac{4T}{200}$. Initial $J_2$ value grid range is the same as the previous scenario, but the grid number becomes 20, not 200. $J_3$ coefficient grid range is initially defined as the minimum value of $-10^{-2}$ and the maximum value of $-10^{-4}$. The grid number for the $J_3$ is 20. Since we estimate $J_2$ and $J_3$ simultaneously, there are 400 possible combinations of $J_2$ and $J_3$ grids.  

\begin{figure}[htbp]
\centering
\includegraphics[width=0.6\textwidth]{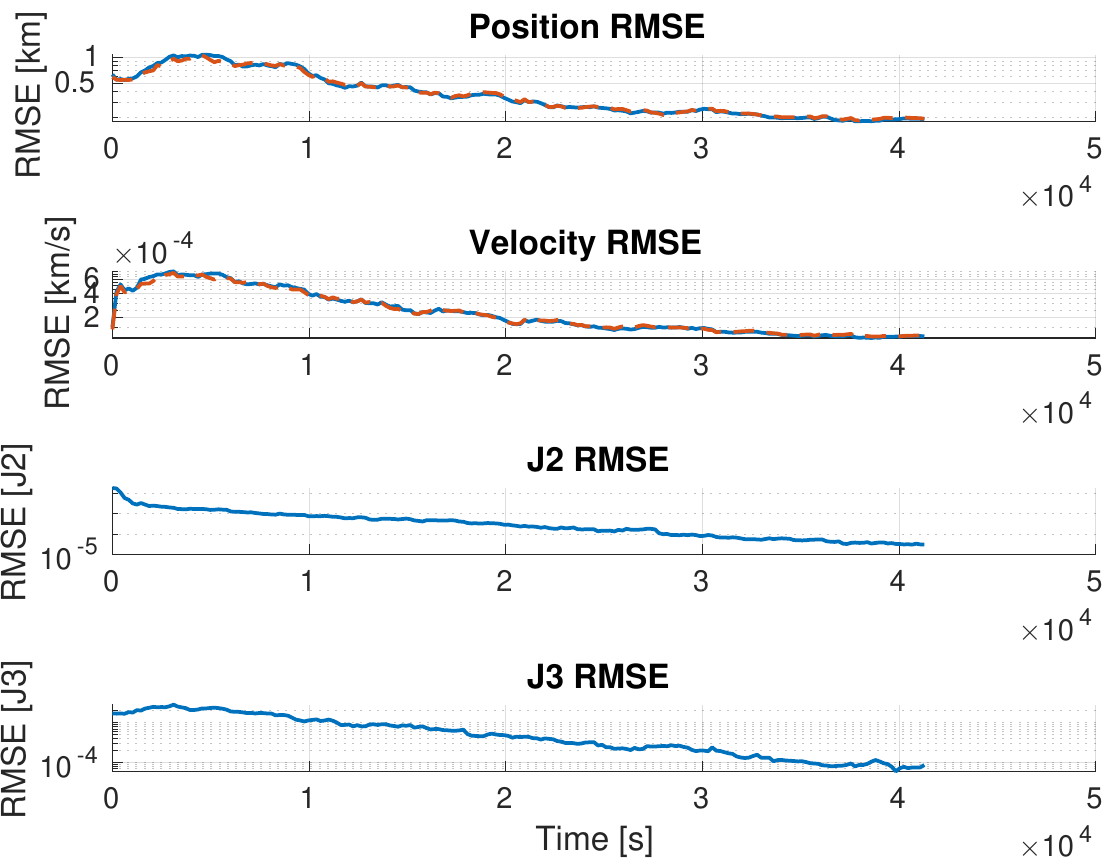}
\caption{RMSE of S/C 1 and S/C 2 for Vesta of $J_2$ and $J_3$ Estimation. Top: Position RMSE for S/C 1 (blue) and S/C 2 (orange). Middle: Velocity RMSE S/C 1 (blue) and S/C 2 (orange). Bottom: $J_2$ and $J_3$ RMSE}
\label{fig: RMSEJ23}
\end{figure}

A 100-run Monte Carlo analysis is utilized to evaluate the statistical performance of the proposed framework for this scenario. As a simulation result, a nominal convergence rate of about 80$\%$ is achieved, which is smaller than in the previous scenario. As a possible reason, by estimating two coefficients, $J_2$ and $J_3$, simultaneously, the framework sometimes estimates different combinations of coefficients that cause similar S/C dynamics. As in previous scenarios, to improve tracking of trend results, outlier results are excluded from the final calculations. Figure~\ref{fig: RMSEJ23} shows the RMSE for each S/C state and the $J_2$ and $J_3$ values.

\begin{figure}[htbp]
\centering
\begin{subfigure}[t]{0.47\textwidth}
    \centering
    \includegraphics[width=\linewidth]{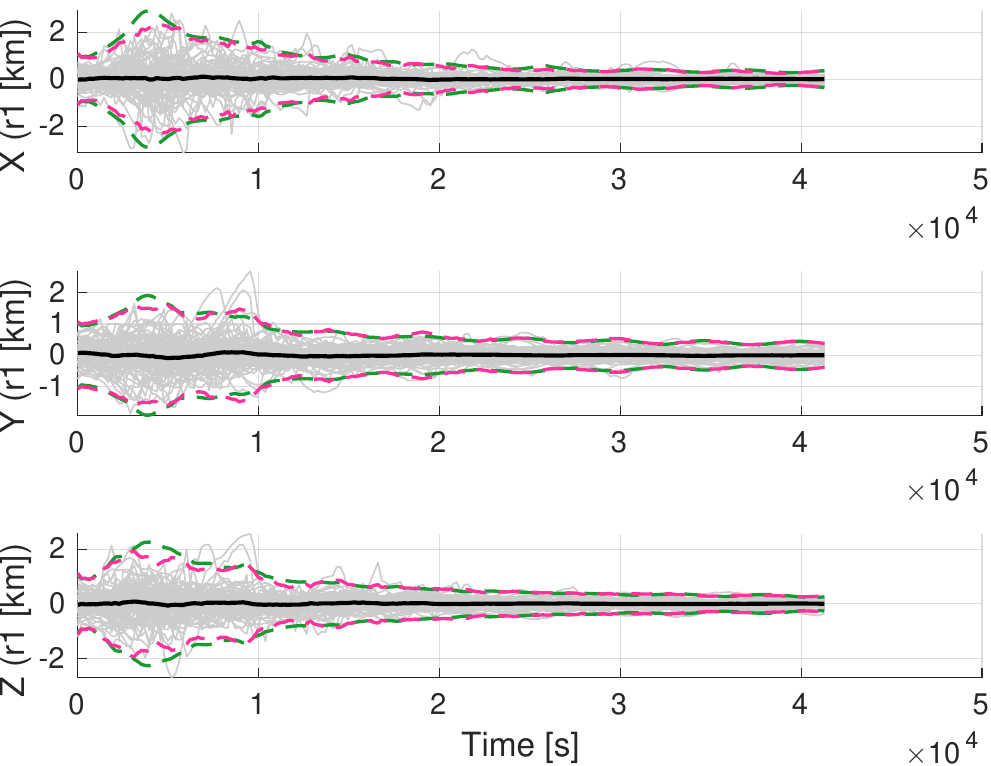}
    \caption{S/C 1 Position}
\end{subfigure}
\hfill 
\begin{subfigure}[t]{0.47\textwidth}
    \centering
    \includegraphics[width=\linewidth]{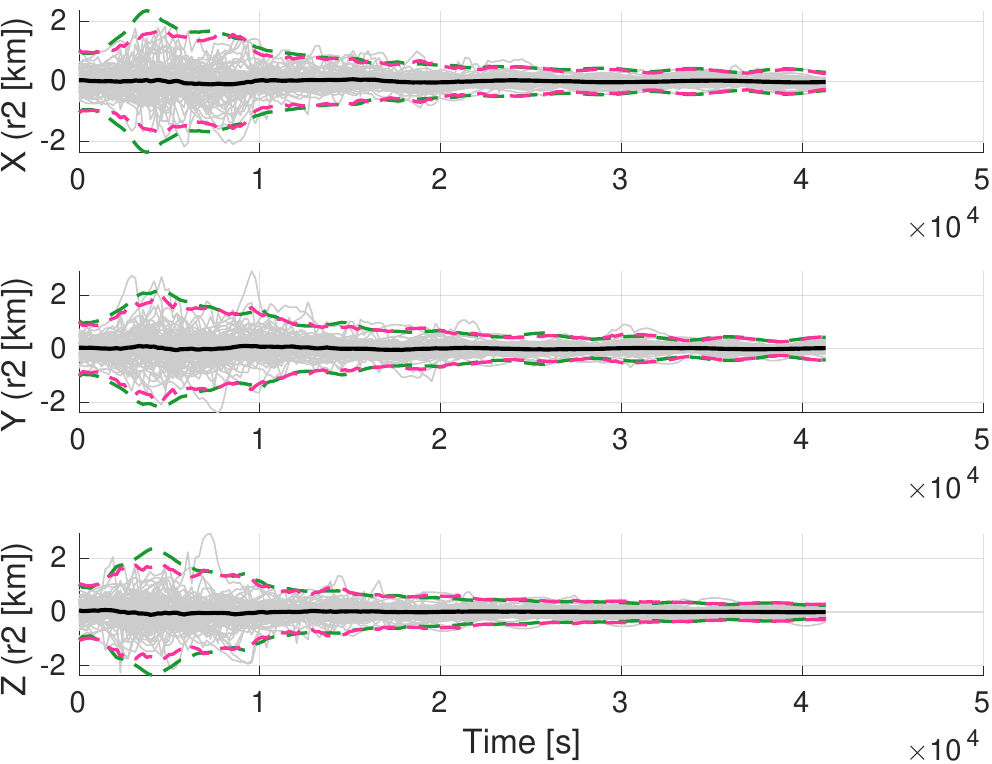}
    \caption{S/C 2 Position}
\end{subfigure}
\\[2ex] 
\begin{subfigure}[t]{0.47\textwidth}
    \centering
    \includegraphics[width=\linewidth]{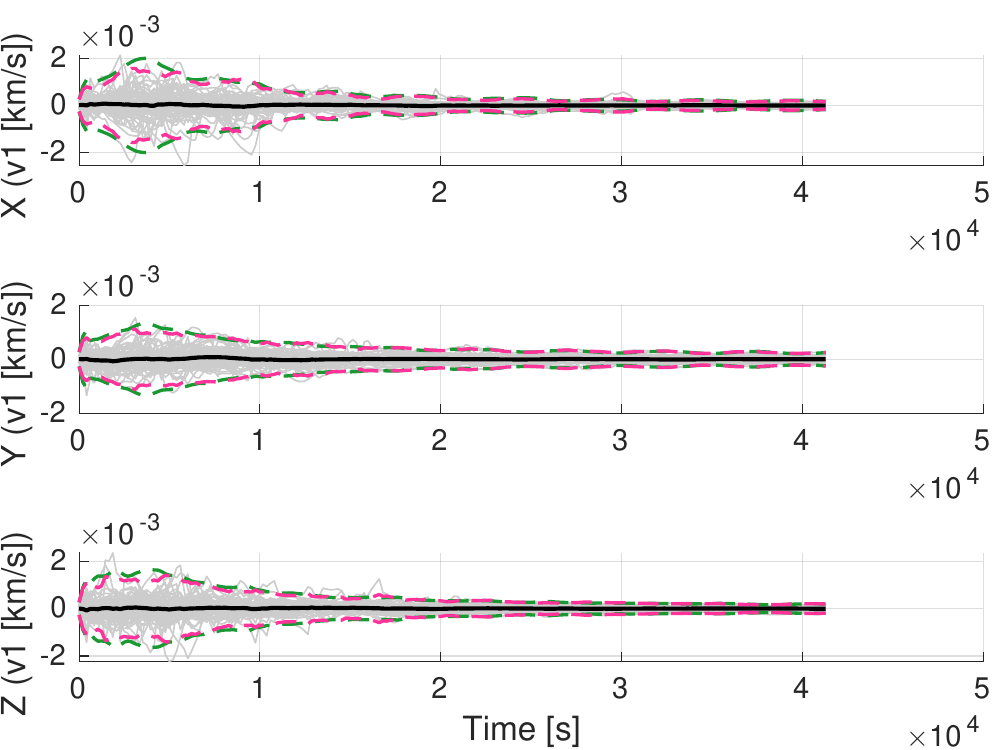}
    \caption{S/C 1 Velocity}
\end{subfigure}
\hfill
\begin{subfigure}[t]{0.47\textwidth}
    \centering
    \includegraphics[width=\linewidth]{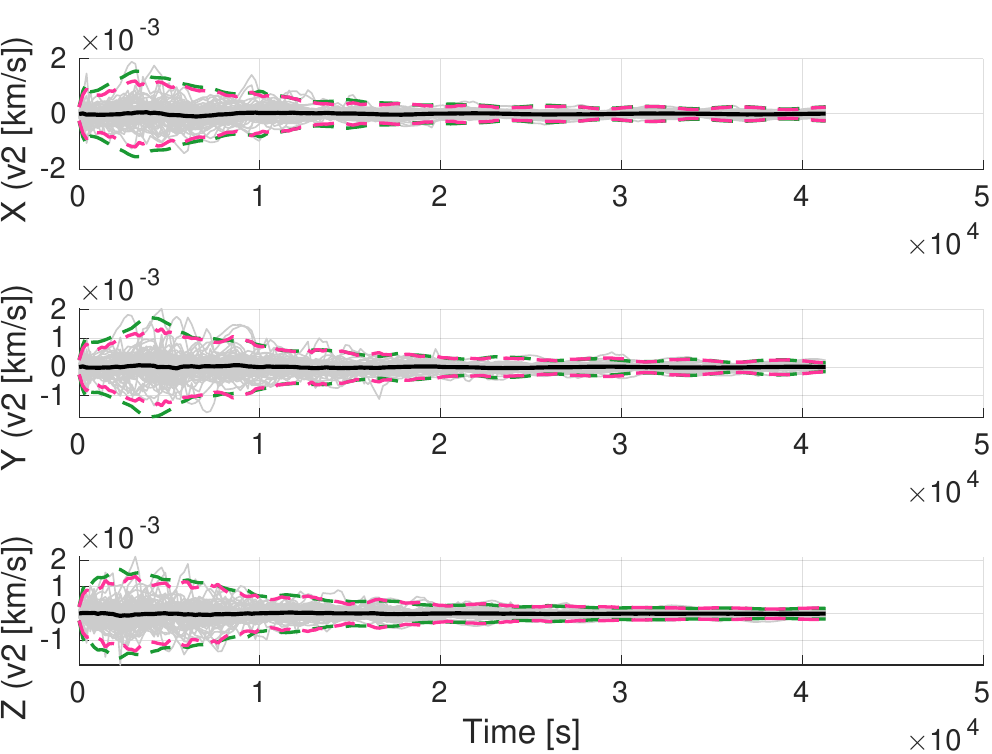}
    \caption{S/C 2 Velocity}
\end{subfigure}
\caption{Two S/Cs Position and Velocity Consistency for Vesta of $J_2$ and $J_3$ Estimation. All plots include estimated $\pm3\sigma$ (green) and empirical $\pm3\sigma$ (pink) bounds across Monte Carlo trials (gray).}
\label{fig: ConcistV23}
\end{figure}

\begin{figure}[htbp]
\centering
\includegraphics[width=0.6\textwidth]{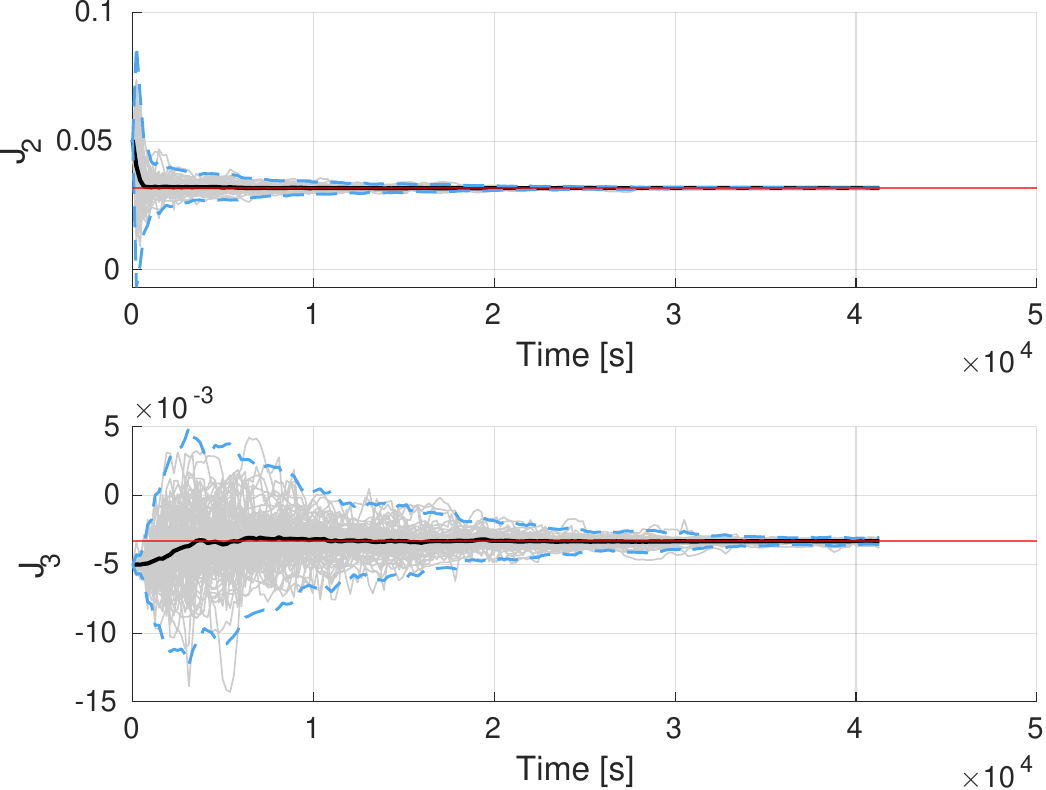}
\caption{$J_2$ and $J_3$ Estimation Consistency for Vesta with $\pm3\sigma$ (blue) bounds}
\label{fig: ConsistJ23}
\end{figure}

Figure~\ref{fig: RMSEJ23} clearly shows the position and velocity RMSEs for each S/C gradually decrease to small values with some oscillation. However, compared with the previous case, it takes much longer for the RMSE to reach a small value, since in this case it is necessary to find the best combination of gravitational perturbations rather than a single parameter. Also, the total grid count is increasing compared with the previous case, while each model's grid count is much smaller. Despite that, the LiAISON estimation architecture overall preserves accurate state and gravitational coefficients estimates for both spacecrafts in orbits around Vesta.
To check the estimation consistency, there are Figure~\ref{fig: ConcistV23} for the state estimation and Figure~\ref{fig: ConsistJ23} for the gravitational coefficient estimation.   

\begin{figure}[htbp]
\centering
\includegraphics[width=0.7\textwidth]{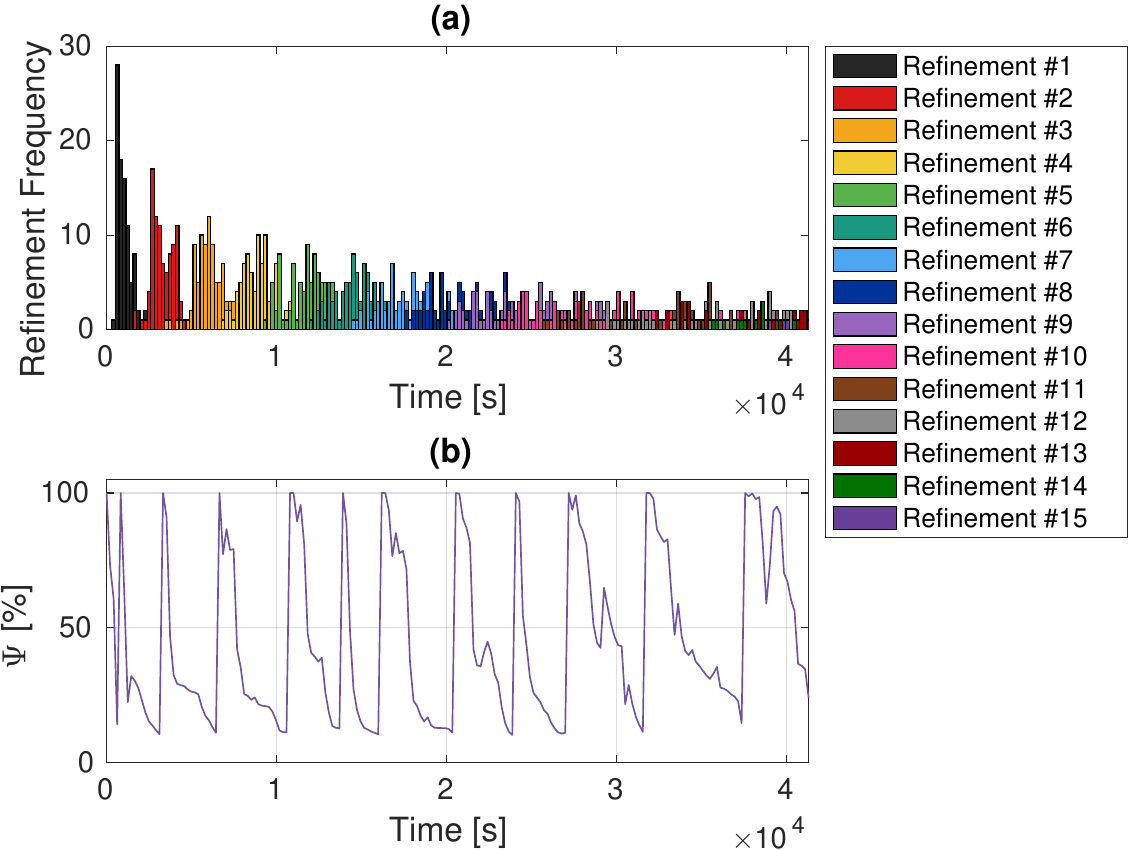}
\caption{Histogram of Refinement Event Times over 100 Monte Carlo Runs (a) and Example Evolution of Hypothesis Diversity Metric $\Psi (t)$ (b) for Vesta of $J_2$ and $J_3$ Estimation}
\label{fig: J2psiV23}
\end{figure}

For the S/C state consistency figure, most estimation errors in each Monte Carlo run are within $\pm3\sigma$ ranges. However, in the $J_2$ estimation case, the standard deviation of the initial simulation is larger. One possible reason is that at the initial time, the estimation framework struggles to find a good coefficient combination and to converge to the true one. For the gravitational coefficient estimation consistency figure, the rate of decrease in $J_3$ errors is much slower than that of $J_2$ errors. However, both the coefficient errors approach zero at the end of the simulation. In fact, the $J_3$ value is about 10 $\%$ of $J_2$, so that the acceleration from the $J_3$ influence is less effective on each S/C dynamics compared with the $J_2$ one. From that point onward, $J_3$ estimation does not perform as well as $ J_2$'s prediction. Although there are some challenges, the state and gravitational coefficient estimation are consistent, as each MC error value converges before the end of the simulation and remains within 3 times the standard deviation.    
Figure~\ref {fig: J2psiV23} represents the behavior of the MMAE $\Psi(t)$ hypothesis-triggering strategy for Vesta in $J_2$ and $J_3$ estimation by using the same format of Figures~\ref {fig: J2psi}. 

From Figure~\ref {fig: J2psiV23} and previous case results, the number of refinements is clearly larger since the refinement number of this scenario is around 15, but the previous case is about 4 times. Unlike the previous simulation case, this scenario requires finding two coefficient values simultaneously, and some incorrect combinations yield state estimation results that differ but are similar. Also, the evolution behavior of $\Psi (t)$ indicates that the result does not converge yet. However, the time spent for $\Psi (t)$ to be less than the critical value is gradually increasing. This proposes that, even if the estimation of a few gravitational coefficients is required, the proposed method works at least in the region around Vesta. 

\section{Conclusion}
This paper evaluates the LiAISON method by estimating the gravitational harmonic coefficients for Ceres and Vesta using the MMAE and UKF framework. Ceres and Vesta are good candidates for simulating the LiAISON approach in near two-body systems because both asteroids have sufficiently strong asymmetric gravity fields, only with $J_2$ influence. The approach combined near two-body dynamics, a basic range and range rate measurement model, the Multi-Model Adaptive Estimation (MMAE) with Unscented Kalman filter (UKF) blocks, and the hypothesis trigger $\Psi(t)$ approach. The numerical results provide sufficient evidence that the LiAISON method is useful for S/C absolute-state estimation on Ceres under the $J_2$  influence condition and on Vesta under the $J_2$ and $J_3$ effect condition. Monte Carlo simulation shows that the LiAISON MMAE with the UKF framework provides accurate and consistent estimates of position, velocity, and the gravitational coefficient. Also, Vesta is more favorable for the LiAISON method because of faster RMSE convergence in position, fewer outliers, and sufficient Performance for $J_2$ and $J_3$ estimation. Although the LiAISON MMAE with the UKF framework estimates reasonable gravitational coefficients, the framework does not converge in the behavior of the hypothesis trigger $\Psi (t)$.               

Future work includes developing a strategy for translating MMAE gravitational harmonic coefficient estimates into spacecraft (S/C) state estimates, treating the coefficients as the state once their values converge sufficiently. Besides, we might need to develop an approach to efficiently and accurately estimate a few coefficients, even when a coefficient is much smaller than the others, such as $J_2$ and $J_3$ for Ceres, and the true and estimated dynamics differ.

\bibliographystyle{AAS_publication}   
\bibliography{bib/refs}   

@article{Choi_2026_Ambiguity,
  title={Ambiguity-Free Geometric Approach to Deep-Space Navigation Using Angles-Only Measurements},
  author={Choi, Sungmoon and Negrete, Aimar and Abdelkhalik, Ossama and Servadio, Simone and Lee, David D},
  journal={Journal of Guidance, Control, and Dynamics},
  pages={1--14},
  year={2026},
  publisher={American Institute of Aeronautics and Astronautics},
  DOI= {10.2514/1.G009699}
}

@INPROCEEDINGS{Stacey_2018_Autonomous,
  author={Stacey, Nathan and D’Amico, Simone},
  booktitle={AAS/AIAA Astrodynamics Specialist Conference}, 
  title={Autonomous swarming for simultaneous navigation and asteroid characterization}, 
  year={2018},
  volume={},
  number={},
  pages={},
  url = {https://slab.sites.stanford.edu/sites/g/files/sbiybj25201/files/media/file/asm_2018_paper_staceydamico_v2.pdf}
}

@article{Lee_2018_Relative,
author = {Lee, Jongwoo AND Park, Sang-Young AND Kang, Dae-Eun},
journal = {Journal of Astronomy and Space Sciences},
publisher = {The Korean Space Science Society},
title = {Relative Navigation with Intermittent Laser-based Measurement for Spacecraft Formation Flying},
year = {2018},
volume = {35},
number = {3},
url = {https://doi.org/10.5140/JASS.2018.35.3.163},
pages = {163-173},
doi = {10.5140/JASS.2018.35.3.163}
}

@INPROCEEDINGS{Michaelson_2024_Particle,
  author={Michaelson, Kristen and Popov, Andrey A. and Zanetti, Renato and DeMars, Kyle J.},
  booktitle={2024 27th International Conference on Information Fusion (FUSION)}, 
  title={Particle Flow with a Continuous Formulation of the Nonlinear Measurement Update}, 
  year={2024},
  volume={},
  number={},
  pages={1-8},
  doi={10.23919/FUSION59988.2024.10706508}
}

@misc{NASA_JPLV,
  author       = {{NASA/JPL Solar System Dynamics}},
  title        = {{JPL Small-Body Database Lookup: 4 Vesta (A807 FA)}},
  howpublished = {\url{https://ssd.jpl.nasa.gov/tools/sbdb_lookup.html#/?sstr=4}},
  year         = {2026},
  note         = {Jet Propulsion Laboratory, California Institute of Technology. Accessed February 27, 2026}
}

@misc{NASA_JPLC,
  author       = {{NASA/JPL Solar System Dynamics}},
  title        = {{JPL Small-Body Database Lookup: 1 Ceres (A801 AA)
}},
  howpublished = {\url{https://ssd.jpl.nasa.gov/tools/sbdb_lookup.html#/?sstr=1}},
  year         = {2026},
  note         = {Jet Propulsion Laboratory, California Institute of Technology. Accessed February 27, 2026}
}

@misc{Ganganath_2026_Compensating,
      title={Compensating Star-Trackers Misalignments with Adaptive Multi-Model Estimation}, 
      author={Ridma Ganganath and Simone Servadio and David Daeyoung Lee},
      year={2026},
      eprint={2601.01130},
      archivePrefix={arXiv},
      primaryClass={eess.SY},
      url={https://arxiv.org/abs/2601.01130}, 
}

@misc{Candan_2025_Adaptive,
      title={Adaptive Relative Pose Estimation Framework with Dual Noise Tuning for Safe Approaching Maneuvers}, 
      author={Batu Candan and Simone Servadio},
      year={2025},
      eprint={2507.16214},
      archivePrefix={arXiv},
      primaryClass={cs.RO},
      url={https://arxiv.org/abs/2507.16214}, 
}

@INPROCEEDINGS{Wan_2000_The,
  author={Wan, E.A. and Van Der Merwe, R.},
  booktitle={Proceedings of the IEEE 2000 Adaptive Systems for Signal Processing, Communications, and Control Symposium (Cat. No.00EX373)}, 
  title={The unscented Kalman filter for nonlinear estimation}, 
  year={2000},
  volume={},
  number={},
  pages={153-158},
  doi={10.1109/ASSPCC.2000.882463}
}

@article{Konopliv_2014_The,
title = {The Vesta gravity field, spin pole and rotation period, landmark positions, and ephemeris from the Dawn tracking and optical data},
author = {A.S. Konopliv and S.W. Asmar and R.S. Park and B.G. Bills and F. Centinello and A.B. Chamberlin and A. Ermakov and R.W. Gaskell and N. Rambaux and C.A. Raymond and C.T. Russell and D.E. Smith and P. Tricarico and M.T. Zuber},
journal = {Icarus},
volume = {240},
pages = {103-117},
year = {2014},
note = {Bright and Dark Materials on Vesta},
issn = {0019-1035},
doi = {https://doi.org/10.1016/j.icarus.2013.09.005}
}

@article{Konopliv_2018_The,
title = {The Ceres gravity field, spin pole, rotation period and orbit from the Dawn radiometric tracking and optical data},
author = {A.S. Konopliv and R.S. Park and A.T. Vaughan and B.G. Bills and S.W. Asmar and A.I. Ermakov and N. Rambaux and C.A. Raymond and J.C. Castillo-Rogez and C.T. Russell and D.E. Smith and M.T. Zuber},
journal = {Icarus},
volume = {299},
pages = {411-429},
year = {2018},
issn = {0019-1035},
doi = {https://doi.org/10.1016/j.icarus.2017.08.005}
}

@book{Battin_1999_An,
  title     = {An Introduction to the Mathematics and Methods of Astrodynamics, Revised Edition},
  publisher = {AIAA},
  author    = {Richard H. Battin},
  year      = {1999}
}

@book{Curtis_2014_Orbital,
  title     = {Orbital Mechanics for Engineering Students, Third Edition},
  publisher = {Elsevier},
  author    = {Howard D. Curtis},
  year      = {2014}
}

@inbook{Fujimoto_2016_Stereoscopic,
author = {Kohei Fujimoto and Nathan Stacey and James M. Turner},
title = {Stereoscopic Image Velocimetry as a Measurement Type For Autonomous Asteroid Gravimetry},
booktitle = {AIAA/AAS Astrodynamics Specialist Conference},
chapter = {},
pages = {},
publisher={AIAA},
doi = {10.2514/6.2016-5566},
year={2016}
}

@article{Turan_2022_Autonomous3,
   title={Autonomous Crosslink Radionavigation for a Lunar CubeSat Mission},
  author={Turan, Erdem and Speretta, Stefano and Gill, Eberhard},
   volume={3},
   ISSN={2673-5075},
   url={http://dx.doi.org/10.3389/frspt.2022.919311},
   DOI={10.3389/frspt.2022.919311},
   journal={Frontiers in Space Technologies},
   publisher={Frontiers Media SA},
   year={2022},
}

@phdthesis{Hill_2007_Autonomous,
  type   = {phd},
  title  = {Autonomous Navigation in Libration Point Orbits},
  school = {University of Colorado},
  author = {Keric A. Hill},
  year   = {2007}
}

@article{Bertone_2018_Impact,
title = {Impact analysis of the transponder time delay on radio-tracking observables},
author = {Stefano Bertone and Christophe {Le Poncin-Lafitte} and Pascal Rosenblatt and Valéry Lainey and Jean-Charles Marty and Marie-Christine Angonin},
journal = {Advances in Space Research},
volume = {61},
number = {1},
pages = {89-96},
year = {2018},
issn = {0273-1177},
doi = {https://doi.org/10.1016/j.asr.2017.09.003}
}

@article{Turan_2022_Autonomous,
title = {Autonomous navigation for deep space small satellites: Scientific and technological advances},
author = {Erdem Turan and Stefano Speretta and Eberhard Gill},
journal = {Acta Astronautica},
volume = {193},
pages = {56-74},
year = {2022},
issn = {0094-5765},
doi = {https://doi.org/10.1016/j.actaastro.2021.12.030}
}

@article{Tieze_2023_BioSentinel,
title = {BioSentinel: A Biological CubeSat for Deep Space Exploration},
author = {Sofia Massaro Tieze and  Lauren C Liddell and Sergio R Santa Maria and  Sharmila Bhattacharya},
journal = {Astrobiologys},
volume = {23},
pages = {631-636},
year = {2023},
doi = {doi.org/10.1089/ast.2019.2068}
}

@article{Krauske_2020_Implementation,
title = {Implementation of commercial Li-ion cells on the MarCO deep space CubeSats},
author = {Frederick C. Krause and Jessica A. Loveland and Marshall C. Smart and Erik J. Brandon and Ratnakumar V. Bugga},
journal = {Journal of Power Sources},
volume = {449},
pages = {227544},
year = {2020},
issn = {0378-7753},
doi = {https://doi.org/10.1016/j.jpowsour.2019.227544},
}

@article{Woellert_2011_CubeSats,
title = {Cubesats: Cost-effective science and technology platforms for emerging and developing nations},
author = {Kirk Woellert and Pascale Ehrenfreund and Antonio J. Ricco and Henry Hertzfeld},
journal = {Advances in Space Research},
volume = {47},
number = {4},
pages = {663-684},
year = {2011},
issn = {0273-1177},
doi = {https://doi.org/10.1016/j.asr.2010.10.009}
}

\end{document}